\documentclass[11pt]{article} 

\usepackage[utf8]{inputenc} 
\usepackage{geometry} 
\usepackage{graphicx} 
\usepackage{booktabs} 
\usepackage{array} 
\usepackage{paralist} 
\usepackage{verbatim} 
\usepackage{subfig} 
\usepackage{fancyhdr} 
\usepackage{sectsty}
\allsectionsfont{\sffamily\mdseries\upshape}
\usepackage[nottoc,notlof,notlot]{tocbibind} 
\usepackage[titles,subfigure]{tocloft}

\usepackage{amsmath,amssymb,epsfig,amsfonts,epsfig,graphicx}
\usepackage{amsmath,amssymb,epsfig,amsfonts,epsfig,graphicx}
\usepackage{epsfig}
\usepackage{amssymb}
\usepackage{amsmath} \usepackage{stmaryrd}
\usepackage{epic} \usepackage{eepic}
\usepackage{latexsym}
\usepackage{fancyhdr}
\usepackage{color}
\usepackage [colorlinks =true, hyperindex =true,citecolor=blue,linkcolor=blue]{ hyperref}
\usepackage[sort&compress]{natbib}
\usepackage[english]{babel}
\usepackage{blindtext}
\usepackage{mathptmx}
\usepackage[scaled=.90]{helvet}
\usepackage{xcolor}
\usepackage{titlesec}

\titleformat{\section}
  {\normalfont\sffamily\bfseries\color{black}}
  {\thesection}{1em}{}

\titleformat{\subsection}
  {\normalfont\sffamily\it\color{black}}
  {\thesubsection}{1em}{}

\newcommand{\beq}{\begin{equation}}
\newcommand{\eeq}{\end{equation}}
\newcommand{\bea}{\begin{eqnarray}}
\newcommand{\eea}{\end{eqnarray}}
\newcommand{\beas}{\begin{eqnarray*}}
\newcommand{\eeas}{\end{eqnarray*}}
\renewcommand{\rho}{\varrho}
\renewcommand{\phi}{\varphi}
\newcommand{\nn}{\nonumber}

\newcommand{\mbs}[1]{\mathbf{#1}}

\newcommand{\crev}{}

 \def\bT{{\mbs{T}}}

 \def\be{{\mbs{e}}}

  \def\bx{{\mbs{x}}}
 
\def\bphi{\hat {\boldsymbol\varphi}_h}
\def\bphin{\hat {\boldsymbol\varphi}_n}
\def\hphi{{\hat \varphi}_h}

\def\bPn{{\mbs{\hat P}_n}} 
 
\def\bP{{\mbs{\hat P}_h}} 
\def\bTh{{\mbs{\hat T}_h}} 
\def\bC{{\hat {\mbs{C}}_h}}
\def\bCn{{\hat {\mbs{C}}_n}}
\def\bCp{{\hat {\mbs{C}}_h^{+}}}

\def\bh{\hat {\mbs{h}}}
\def\bk{\hat {\mbs{k}}}

\makeatletter
\def\ps@pprintTitle{%
\let\@oddhead\@empty
\let\@evenhead\@empty
\let\@oddfoot\@empty
\let\@evenfoot\@oddfoot
}
\makeatother

\title{On the correspondence between 2D force networks, tensegrity structures and polyhedral stress functions}

\author{Fernando Fraternali and Gerardo Carpentieri \bigskip \\
Department of Civil Engineering, University of Salerno \smallskip \\
Via Ponte Don Melillo, 84084, Fisciano (SA), Italy}

\date{} 

\begin{document}
\maketitle

\renewcommand{\abstractname}{\textsc{Abstract}}

\begin{abstract}

We formulate and discuss the relationship between polyhedral stress functions and internally self-equilibrated frameworks in 2D, and a
two-mesh technique for the prediction of the stress field associated with such systems.  
We generalize classical results concerned with smooth Airy stress functions to polyhedral functions associated with arbitrary triangulations of a simply-connected domain.
We also formulate a regularization technique that smoothly projects the stress function corresponding to an unstructured force network over a structured triangulation. 
The paper includes numerical examples dealing with a benchmark problem of plane elasticity, and the stress fields associated with tensegrity models of a cantilever beam and an elliptical membrane.

\end{abstract}


\let\thefootnote\relax\footnote{Email addresses: f.fraternali@unisa.it (F. Fraternali), gcarpentieri@unisa.it (G. Carpentieri)}

\section{INTRODUCTION}
\label{secintro}

Over recent years, several researchers have focused their attention on the modeling of continuous media such as plates, walls, membranes, vaults and domes with `equivalent' truss structures
(refer, e.g., to \cite{Dwy:1999, LSM1, LSM2, DaviniParoni:2003, Kil:2005, Blo:2007, Blo:2009, Micheletti:2008, LSM4, LSM5, Blo:2011, Angelillo:2012, Desbrun:2013, DeGoes:2013}, and therein references).
Numerious up-to-date contributions to such a longly debated topic of structural mechanics deal with `non-conforming' or `mixed' finite element methods, also referred to as Lumped Stress Methods (LSMs) \cite{LSM1, LSM2, LSM4, LSM5}; the so-called Thrust Network Analysis (TNA), reciprocal force diagrams and limit analysis approaches \cite{Dwy:1999, Blo:2007, Micheletti:2008, Blo:2009, Blo:2011,Angelillo:2012}, as well as Discrete Exterior Calclus (DEC) \cite{Desbrun:2013, DeGoes:2013}. A common trait of the above methods consists of looking at the approximating truss structure as the support of uniaxial singular (or lumped) stresses, which approximate the stress field of the background medium. Studies regarding the convergence of a singular discrete stress network to its continuum limit have been carried out through Gamma-Convergence \cite{DaviniParoni:2003}, and mixed finite element methods \cite{LSM3}.
Particular attention has been devoted to masonry structures described through the no-tension constitutive model \cite{Giaquinta:1985}, since for such structures the singular stress approach allows one to linearize the no-tension constraint, and to make use of form-finding approaches based on convex-hull techniques and weighted Delaunay triangulations  \cite{Dwy:1999, Blo:2007, Blo:2009, LSM4, LSM5, Blo:2011, Angelillo:2012, DeGoes:2013}.
Remarkable is the use of polyhedral Airy stress functions in 2D elasticity problems, and Pucher's approaches to the membrane theory of shells \cite{Mansfield:1964}, which leads to an effective characterization of internally self-equilibrated frameworks associated with simply connected domains
\cite{LSM1, LSM2, LSM3,LSM4, LSM5}.

Force networks are also employed  within`atomistic' models and discrete-continuum approaches to mechanical systems, to represent the state of stress of solids, fluids and biomechanical systems.
Coupled discrete-continuum approaches combine force networks and continuous stress fields (refer, e.g., to \cite{Miller2009} for an extensive review), in order to circumvent scaling limitations of fully atomistic models, which are particularly suited to describe small process zones (interested, e.g., by dislocation and fracture nucleation, nanoindentation, marked atomic rearrangements, etc.). 
Areas of research involving discrete models of mechanical systems include
bio- and nano-structures \cite{Huang2004, Tu:2008, Fra2:2010, CNTIDENT, BFRAD11,FraMar:2012, FraJMPS:2012,  FraMar2:2012}; 
tensegrity models of engineering and biological systems \cite{Ske:2002, Ver:2005, Mof:2006, Ske:2010, Fra:2012};  
structural optimization and form-finding methods \cite{LSM4, LSM5, Lin:2010, Ohmori:2011, Wang:2010, Baldassini:2010},
and strut and tie models of discontinuous regions in reinforced-concrete structures \cite{Schlaich}, just to name a few examples.
Key aspects of scale-bridging approaches to discrete systems regard the estimation of the  Cauchy stress at the meso-scale, to be carried out via statistical mechanics, variational approaches, and/or homogenization methods. 
Several discrete (or `microscopic') definitions of the Cauchy stress have been proposed in the literature, such as, e.g., the \textit{virial stress}, 
the \textit{Tsai traction} and the \textit{Hardy stress} (cf.,e.g., \cite{Shen2004, Admal2010}, and therein references). 
Different studies have highlighted issues related to the kinetic terms of such stress definitions \cite{Shen2004}, and spatial fluctuations of the discrete stress (cf. Sect. 6 of  \cite{Admal2010}). 

The present work deals with the correspondence between polyhedral (Airy) stress functions, internally self-equilibrated force networks, and discrete notions of the Cauchy stress in two-dimensions.
We extend previous research on such topics \cite{LSM1, LSM2, LSM3, LSM4, LSM5}, on examining two new subjects: $(i)$ the computation of the Airy stress function associated with a given, internally self-equilibrated framework; $(ii)$ the formulation of convergent estimates of the Cauchy stress associated with unstructured force networks. Our previous studies in this field were instead focused on the derivation of force networks from a given polyhedral stress function (inverse problem with respect to $(i)$, cf. \cite{LSM1, LSM2, LSM4, LSM5}), and the convergence of stress measures associated with structured force networks \cite{LSM2, LSM3}. 
By examining a simply connected domain in two dimensions, we here develop and discuss an algebraic equation relating polyhedral stress functions and internally self-equilibrated frameworks associated with arbitrary triangulations. Further on, we formulate a regularization technique that is devoted to generate a convergent notion of the Cauchy stress of the discrete system in the continuum limit.
The remainder of the paper is organized as follows.
We begin by framing the correspondence between force networks and polyhedral stress functions in Sect. \ref{airy}.
Next, we formulate a two-mesh approach to the Cauchy stress associated with an unstructured, internally self-equilibrated framework (Sect. \ref{stress}). We illustrate the potential of the proposed approach through a convergence study focused on a benchmark problem of linear elasticity (Sect. \ref{Flamant}); and the state of stress associated with flat and curved tensegrity structures (Sects. \ref{tensegrity} and \ref{elliptical dome}).
We end summarizing the main results and future directions of the present work in Sect. \ref{conclusions}.

\section{INTERNALLY SELF-EQUILIBRATED FRAMEWORKS AND POLYHEDRAL STRESS FUNCTIONS}
\label{airy}

Throughout the paper, we refer to a triangulation $\Pi_h$ of a polygonal and simply-connected domain $\Omega$ of the two-dimensional Euclidean space, which shows $M$ non-degenerate triangles $\Omega_1, ..., \Omega_M$
and features the following {size}:  $h = \mbox{sup}_{m\in\{1,...,M\}} \{diam (\Omega_m)\}$.
We name \textit{`physical'} the edges of $\Pi_h$ that do \textit{not} belong to the boundary of $\Omega$.

\begin{figure*}[htbp]
  \begin{center}
    \epsfig{file=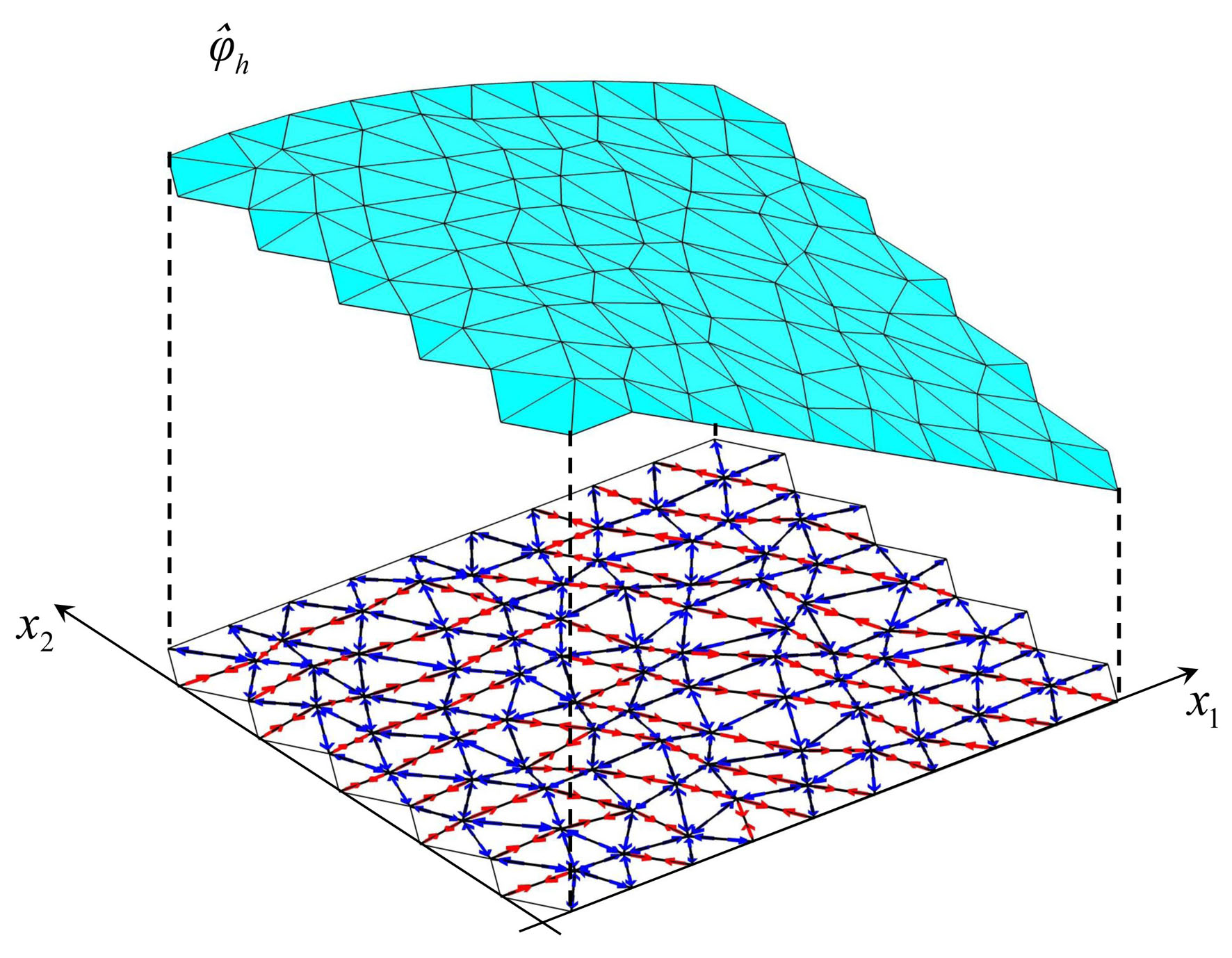,width=12cm} 
   \caption{(Color in the online version). Illustration of a triangulated force network and the associated polyhedral stress function $\hphi$ (red: tensile forces, blue: compressive forces).}
    \label{fig:Pih}
 \end{center}
\end{figure*}

\begin{figure*}[hb]
\unitlength1cm
\begin{picture}(12.5,6.0)
\put(1.5,0){\psfig{figure=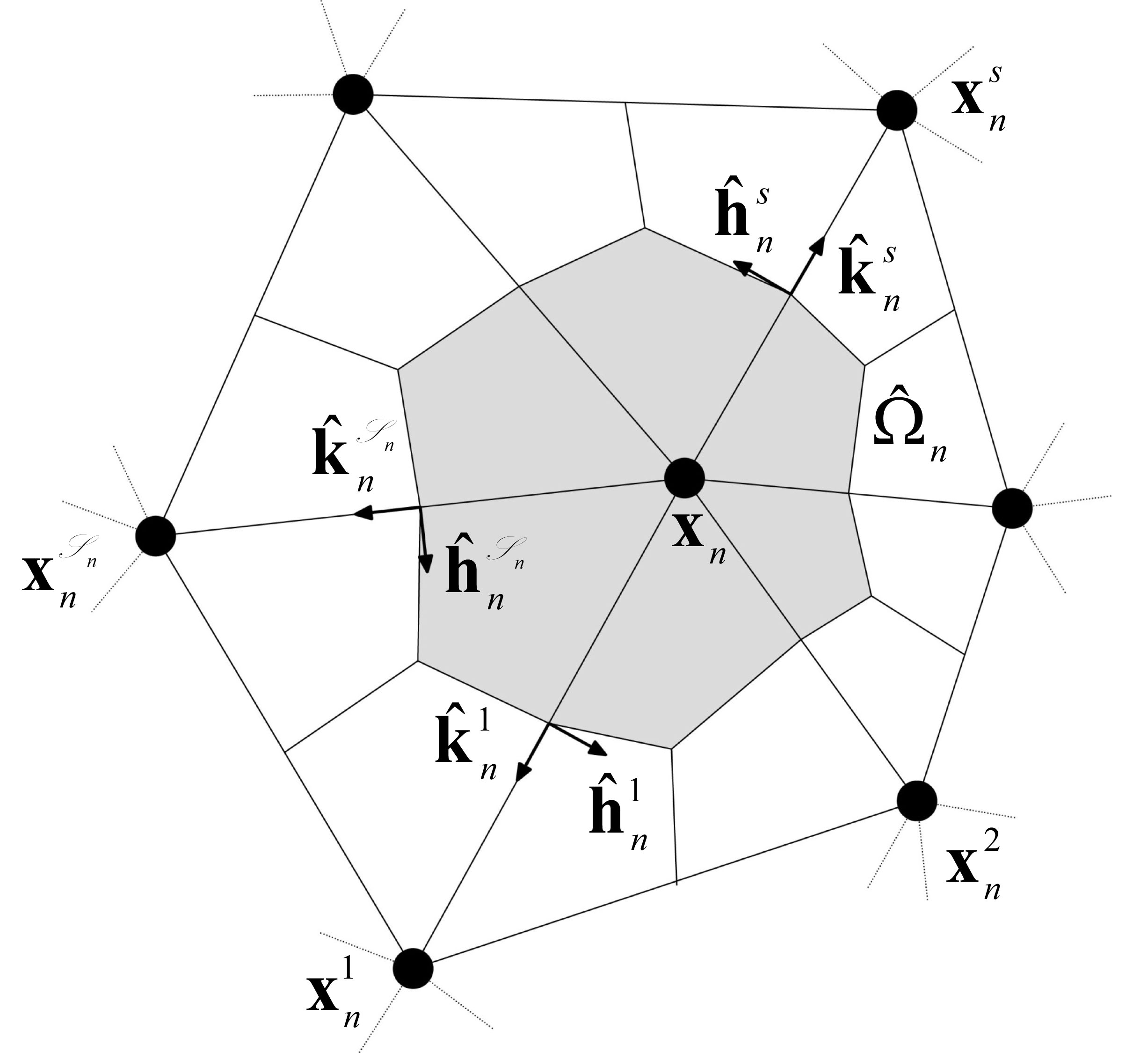,height=6.0cm}}
\put(9.5,0){\psfig{figure=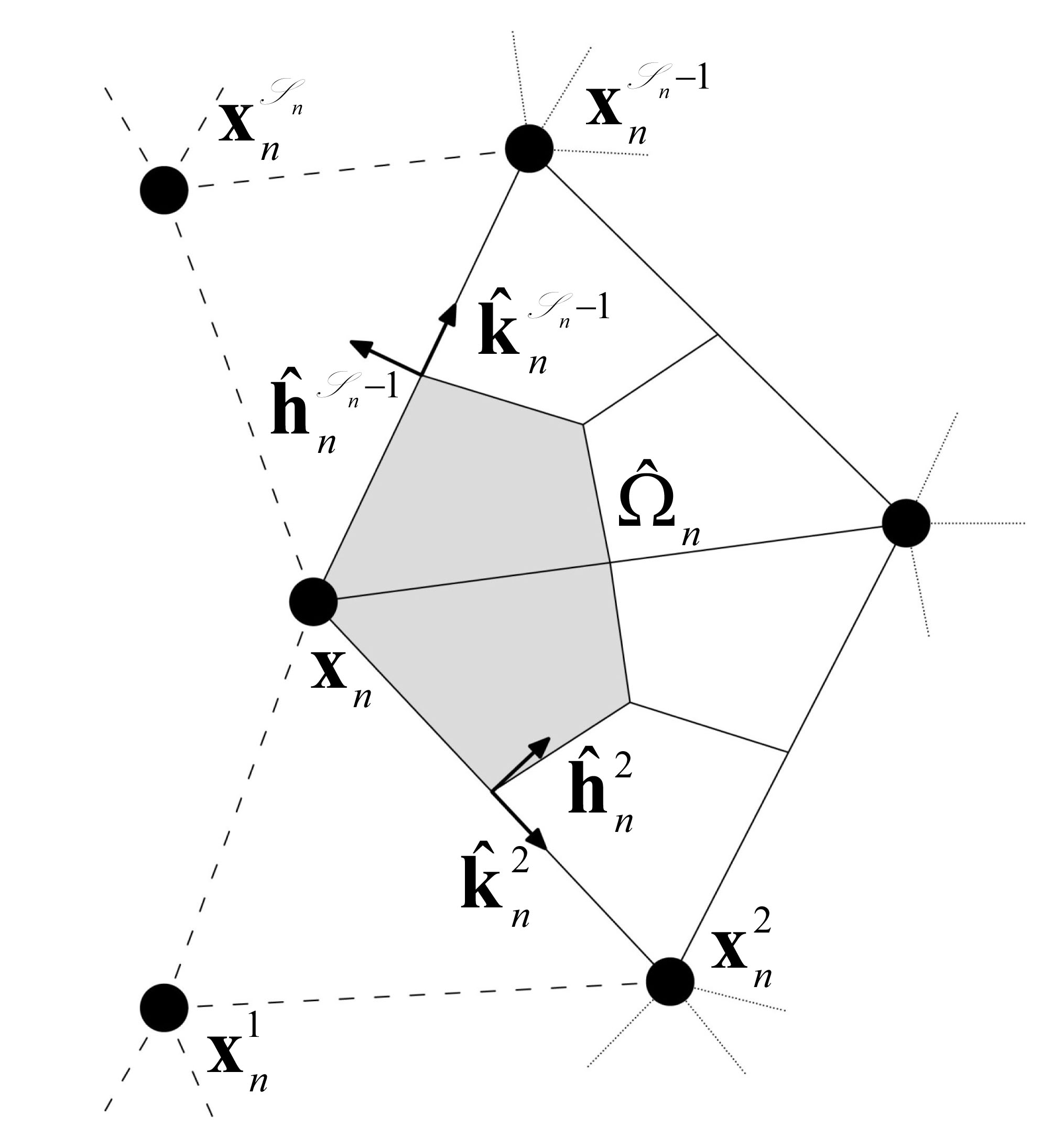,height=6.0cm}}
\end{picture}
\caption{Details of an inner node (left) and a boundary node (right) of $\Pi_h$.}
\label{fig:nodes}
\end{figure*}

\subsection{Internally self-equilibrated framework associated with a given polyhedral stress function}
\label{airy-forces}

Let us introduce Cartesian coordinates $x_1$ and $x_2$ in the plane of $\Omega$ and the polyhedral function defined as follows

\bea
\hphi (\bx) & = &  \sum_{n=1}^{N} {\hat \varphi}_n \ g_n(\bx)
\label{rh}
\eea 

\noindent where $\bx= [ x_1, x_2] ^T$; $N$ is the total number of nodes of the triangulation $\Pi_h$; 
${\hat \varphi}_n$ is the value taken by $\hphi$ at the node $\bx_n$;
and $g_n$ is the piecewise linear basis function associated with such a node (`umbrella' basis function). 
{\crev{We agree to denote the coordination number of $\bx_n$ by ${\cal S}_n$, and
the edges attached to such a node by 
$\Gamma_n^1, ..., \Gamma_n^{{\cal S}_n}$. The unit vectors perpendicular and tangent to $\Gamma_n^1, ..., \Gamma_n^{{\cal S}_n}$  will be hereafter indicated by 
${\bh}_n^1 ,..., {\bh}_n^{{\cal S}_n}$, and ${\bk}_n^1 ,..., {\bk}_n^{{\cal S}_n}$, respectively (Fig. \ref{fig:nodes}).
}}
By interpreting $\hphi$  as a \textit{generalized (Airy) stress function}, we associate a set of $N_{\Gamma}$ forces with such a function, where  $N_{\Gamma}$ indicates the total number of physical edges  of $\Pi_h$.
The generic of such forces is is given by  
\bea
P_n^s & = &  \left[\left[  \nabla \hphi  \cdot {\bh}_n^s   \right] \right]_n^s
\label{Pns}
\eea 

\noindent where $\left[\left[ \nabla \hphi \right] \right]_n^s$ indicates
the jump of the gradient of $\hphi$ across the edge $\Gamma_n^s$ \cite{LSM2, LSM3}. 
The gradient $\nabla \hphi$  is computed as follows over the generic triangle $\bx_n$, $\bx_n^s$, $\bx_n^{t}$  (refer, e.g., to \cite{Glowinski:1973})

\bea
\nabla \hphi & =  & \frac{1}{2 A} \ 
\left[ \begin{array}{l} 
\{ {\hat \varphi}_n (\bx_n^s - \bx_n^t)   +   
{\hat \varphi}_n^s (\bx_n^t - \bx_n) + 
{\hat \varphi}_n^t  (\bx_n - \bx_n^s)  \}  \cdot {\hat \be}_2 \\ 
\{ {\hat \varphi}_n (\bx_n^t - \bx_n^s)   +   
{\hat \varphi}_n^s (\bx_n - \bx_n^t)  + 
{\hat \varphi}_n^t  (\bx_n^s - \bx_n) \} \cdot {\hat \be}_1
\end{array}\right ]
\label{gradphi}
\eea 

\noindent where $A$ is the area of the above triangle, and ${\hat \be}_{\alpha}$ is the unit vector in the direction of the $x_{\alpha}$-axis. 
Equation (\ref{Pns}) shows that the forces $P_n^s$ are associated with the \textit{`folds'} of the graph of $\hphi$.
In particular, \textit{convex folds} of $\hphi$ correspond with \textit {tensile forces}, 
while \textit{concave folds} correspond with \textit {compressive forces} (Fig. \ref{fig:Pih}).
It is useful to recast  (\ref{Pns}) in matrix form, by proceeding as follows.
Let us sort the $\bx_n^1,..., \bx_n^{{\cal S}_n}$ nodes connected  to $\bx_n$ in counterclockwise order, as shown in Fig. \ref{fig:nodes}, and denote the values taken by $\hphi$ at such nodes by 
${\hat \varphi}_n^1,..., {\hat \varphi}_n^{{\cal S}_n}$, respectively.
Said ${\cal P}_n \le {\cal S}_n $ the number of physical edges attached to $\bx_n$, we collect the forces associated with such a node into the ${\cal P}_n$-dimensional vector $\bPn = [ P_n^1,..., P_n^{{\cal P}_n} ]^T$, and the values of ${\hat \varphi}$ at $\bx_n^1,..., \bx_n^{{\cal S}_n}$ and $\bx_n$ into the $({\cal S}^{'}_n = {\cal S}_n + 1)$-dimensional vector $\bphin = [ {\hat \varphi}_n^1,..., {\hat \varphi}_n^{{\cal S}_n}, {\hat \varphi_n} ]^T$.
Straightforward calculations show that the substitution of (\ref{gradphi}) into (\ref{Pns}) leads to the following algebraic equation

\bea
\bPn & = & \bCn \ \bphin
\label{Pnvec}
\eea 

\noindent where $ \bCn$ is the ${\cal P}_n  \times {\cal S}^{'}_n $ matrix defined through

\begin{align}
\label{Cn}
(C_n)_{jk} &= \left\{\! \begin{array}{ll}
a \ = \   { {\bh_n^{j'''}} \cdot  {\bh_n^{j'}} } / {(\ell_n^{j'} ({\bh_n^{j'''}} \cdot  {\bk_n^{j'}}) )} & \\
& \\
\ \ \ \ \ \ - \  
    { {\bh_n^{j''}} \cdot  {\bh_n^{j'}} } / {(\ell_n^{j'} ({\bh_n^{j''}} \cdot  {\bk_n^{j'}}) )} , 
& k = j',\\
& \\
b \ = \  - { {\bh_n^{j'}} \cdot  {\bh_n^{j'}} } / {(\ell_n^{j''} ({\bh_n^{j'}} \cdot  {\bk_n^{j''}}) )} , 
&  k = j'',\\
& \\
c \ = \   { {\bh_n^{j'}} \cdot  {\bh_n^{j'}} } / {(\ell_n^{j'''} ({\bh_n^{j'}} \cdot  {\bk_n^{j'''}}) )} , 
& k = j''',\\
& \\
d \  = \ - a - b - c , 
& k = {\cal S}_n^{'},\\
& \\
0 , 
&\mbox{otherwise,}
\end{array}\right.
\end{align}

\noindent In (\ref{Cn}), $\ell_n^s$ denotes the length of $\Gamma_n^s$, and it results

\beq
\label{jdef}
\begin{array}{llll}
\mbox{\textit{inner node}} & & & \mbox{\textit{boundary node}} \\
j'=j & & & j'=j+1 \\
\mbox{if  } j'>1 \mbox{  then  } j'' = j'-1, \mbox{  else  } j'' ={\cal S}_n & & &  j''=j'-1\\
\mbox{if  } j'<{\cal S}_n \mbox{  then  } j''' = j'+1, \mbox{  else  } j''' =1 & & & j'''=j'+1 
\end{array}
\eeq

By using standard matrix assembling techniques, we finally obtain the following `global' equation

\bea
\bP & = & \bC \ \bphi
\label{Pvec}
\eea 

\noindent which relates the vector  $\bP$ collecting all the forces $P_n^s$ to the  vector $\bphi$ collecting all the nodal values of $\hphi$. In (\ref{Pvec}), $\bC$ is the $N_{\Gamma} \times N$ matrix obtained by assembling the nodal matrices  (\ref{Cn}).
It can be shown \citep{DeGoes:2013} that the forces $\bP$ computed through (\ref{Pvec}) automatically satisfy the equilibrium equations of the internal nodes of $\Pi_h$ with zero external forces, for any  given $\bphi \in \mathbb{R}^N$. This implies that  $\bP$ and the graph structure associated with $\Pi_h$  form an \textit{internally self-equilibrated framework} \citep{DeGuz:2006, Micheletti:2008}. 

\subsection{Polyhedral stress function associated with a given, internally self-equilibrated framework}
\label{forces-airy}

\begin{figure*}[htbp]
  \begin{center}
    \epsfig{file=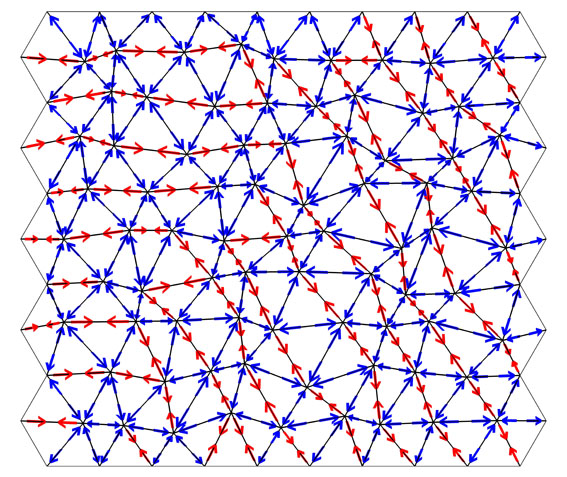,width=10cm} 
   \caption{(Color in the online version). 2D view of the force network in Fig. \ref{fig:Pih} (red: tensile forces, blue: compressive forces).}
    \label{fig:Pih2}
  \end{center}
\end{figure*}

We now pass to examine the problem of finding a polyhedral stress function $\hphi$ associated with a given, internally self-equilibrated framework $\bP$ in two-dimensions. The latter may arise e.g. from pair-interactions in a particle system \cite{Admal2010}, or a lumped stress/tensegrity approach to the  equilibrium problem of a continuous medium \cite{LSM2, Ske:2010}. As anticipated, we assume that $\bP$ is associated with the (physical)  edges of a planar (non-degenerate) triangulation $\Pi_h$ of simply-connected domain $\Omega$. 
 It is clear that the current problem is related to the inversion of the linear system of algebraic equations (\ref{Pvec}).
Let us refer to the illustrative example represented in Fig. \ref{fig:Pih2}, which shows a triangulated force network with a total of $N=115$ nodes;
77 inner nodes; and 266 physical edges. 
We have observed in the previous section that the forces $\bP$  computed through (\ref{Pvec}) satisfy the equilibrium equations of the inner nodes of $\Pi_h$ (with zero applied forces), for any given $\bphi \in \mathbb{R}^N$. This proves that the rank of $\bC$ is equal to 112 ($r=\mbox{rank}(\bC) = 266 - 2\times 77 = 112$), and that the nullity of the same matrix is equal to 3 ($n=\mbox{nullity}(\bC) = 115 -  112 = 3$, cf., e.g., \cite{Strang}),  in the case under examination. Given an arbitrary  internally self-equilibrated force network $\bP \in \mathbb{R}^r$, we therefore conclude the following: $(i)$ the linear system (\ref{Pvec}) actually admits  solutions $\bphi \in \mathbb{R}^N$; $(ii)$ such solutions are determined up to three arbitrary constants; $(iii)$ two solutions differ by linear functions associated with zero axial forces along the edges of $\Pi_h$.  It is not difficult to realize that  the above results $(i)$, $(ii)$ and $(iii)$, which generalize analogous ones concerned with smooth Airy functions \cite{Gurtin}, can be extended to arbitrary triangulations of simply-connected domains. 
Consider, e.g., that the insertion of an additional (inner) node into the triangulation in Fig. \ref{fig:Pih2} leads to a new triangulation carrying 116 nodes; 269 forces: and $2\times 28=156$ equilibrium constraints ($\mbox{rank}(\bC) = 269 - 156 = 113$). 
It is easily shown that such an insertion leaves the nullity of $\bC$ equal to 3.
The indeterminacy of system (\ref{Pvec})  can be resolved by prescribing $\hphi$ at three non-collinear nodes of $\Pi_h$ (e.g., prescribing the values of $\hphi$ at the vertices of a given triangle). A particular solution of (\ref{Pvec}) is given by

\bea
\bphi & = & \bCp \  \bP
\label{Pinv}
\eea 

\noindent where $\bCp$ denotes the Moore-Penrose inverse of $\bC$.

\section{STRESS FIELD ASSOCIATED WITH AN INTERNALLY SELF-EQUILIBRATED FRAMEWORK}
\label{stress}

It is not difficult to realize that  a scale bridging approach to the stress field associated with a self-equilibrated force network $\bP$  can be obtained by introducing a suitable regularization of the corresponding stress function $\hphi$.
Consider, indeed, that the stress field associated with a smooth Airy stress function $\varphi_0$ corresponds with the hessian of $\varphi_0$ (under a suitable rotation transformation, see, e.g., \cite{Green:2002,Gurtin}), i.e. the second-order tensor with Cartesian components $\partial \varphi_0 / \partial x_{\alpha} \partial x_{\beta}$ ($\alpha, \beta = 1,2$).
Since the second-order derivatives of a polyhedral function $\hphi$ exist only in the distributional sense, the definition of a stress field associated with $\hphi$ calls for the introduction of a generalized notion of the hessian of  such a function \cite{DaviniParoni:2003, LSM3}.
A convergent stress measure has been defined in \cite{LSM3}, on considering sequences of polyhedral stress functions associated with structured triangulations. The latter match the ${\cal P}_{\Sigma}$ property defined in Sect. 5 of \cite{LSM3}, and consist, e.g., of triangulations associated with rectangular or hexagonal Bravais lattices (cf. Figs. 2 and 3 of \cite{LSM3}).
Let us define a 'dual mesh' ${\hat \Pi}_h$ of $\Omega$, 
which is formed by polygons connecting the barycenters of the triangles attached to the generic node $\bx_n$ to the mid-points of the edges $\Gamma_n^1, ..., \Gamma_n^{{\cal S}_n}$ ('barycentric'  dual mesh, cf. Fig. \ref{fig:Pih}). The stress measure defined in
\cite{LSM3} is a piecewise constant stress field $\bTh$ over ${\hat \Pi}_h$, which takes the following value in correspondence with the generic dual cell ${\hat \Omega}_n$

\bea
\bTh(n) & = & \frac{1}{|{\hat \Omega}_n|} \sum_{j=1}^{{\cal P}_n} \frac{\ell_n^{j'}}{2} 
P_n^{j'}
 \bk_n^{j'} \otimes \bk_n^{j'} 
\label{Thn}
\eea 

\begin{figure*}[htbp]
  \begin{center} 
    \epsfig{file=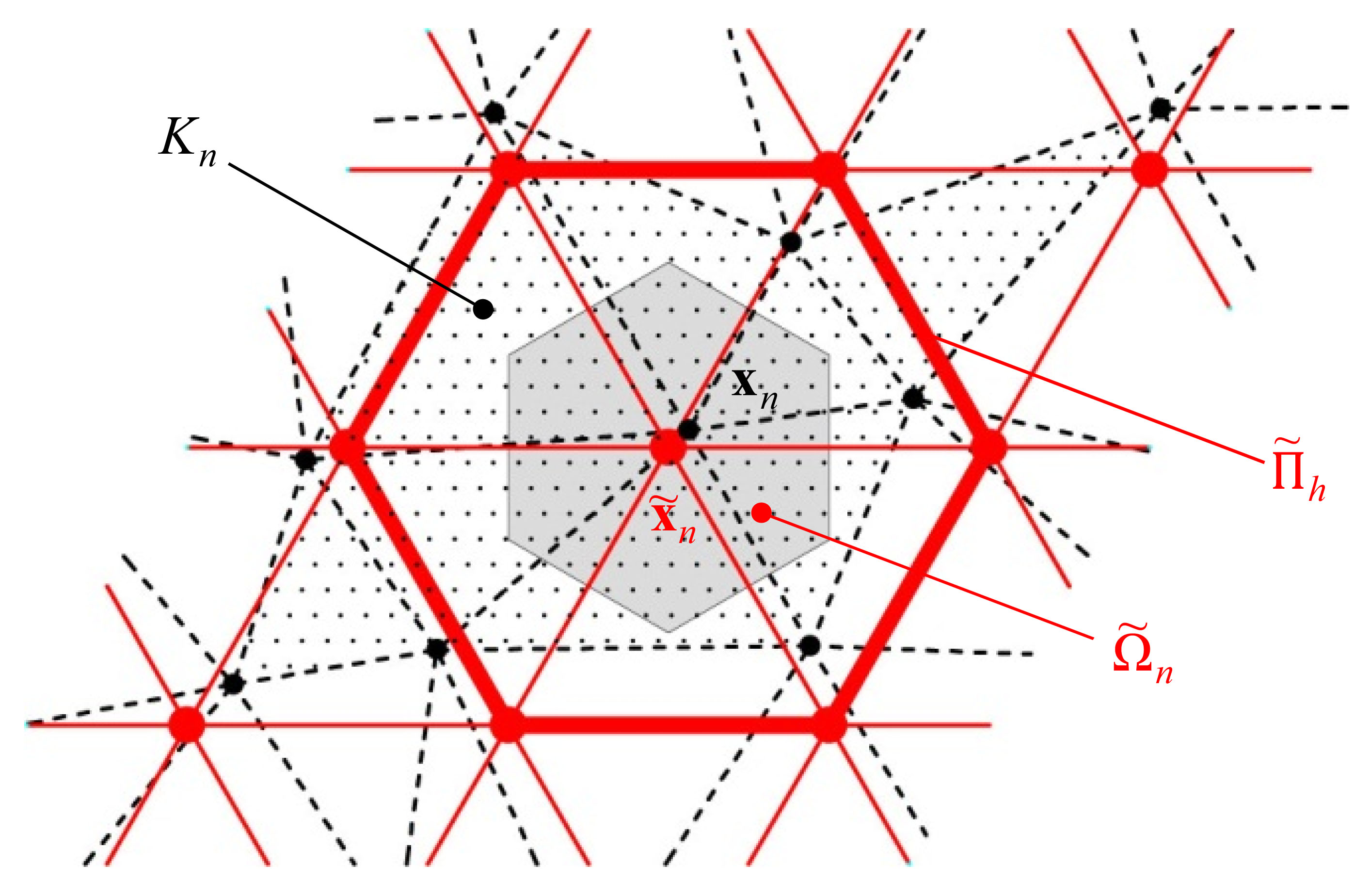,width=14cm} 
   \caption{Illustration of $K_n$ and ${\tilde \Omega}_n$.}
    \label{fig:Kn}
  \end{center}
\end{figure*}

\noindent Here, $|{\hat \Omega}_n|$ denotes the  area of ${\hat \Omega}_n$, and $j'$ is defined as in (\ref{jdef}).
Under the assumption that $\Pi_h$ is a structured triangulation, it has been shown in \cite{LSM3}  that the discrete stress (\ref{Thn}) strongly converges to the stress field associated with the limiting stress function, as the mesh size approaches zero (cf. Lemma 2 of \cite{LSM3}). It is worth observing that $\bTh(n)$ is obtained by looking at the quantity $P_n^{s} \bk_n^{s} \otimes \bk_n^{s}$ as a `lumped stress tensor' acting in correspondence with the edge $\Gamma_n^s$, and that  Eqn. (\ref{Thn}) spatially averages the lumped stress tensors competing to $\bx_n$, over the corresponding dual cell ${\hat \Omega}_n$ (averaging domain).
We also note that the stress measure
(\ref{Thn}) corresponds with the virial stress of statistical mechanics at zero temperature (cf. \cite{Admal2010}, Sect. 2.2 and Appendix A). 
Unfortunately, the error estimate given in Lemma 2 of \cite{LSM3} does not cover {\crev{{unstructured triangulations}}}, as we already noticed. 
We hereafter handle the case of an unstructured polyhedral stress function $\hphi$ by employing the regularization procedure formulated in \cite{Fracur:2013} to predict the curvatures of polyhedral surfaces. Let us consider  
{\crev{an arbitrary vertex $\bx_n$ of $\hphi$, and a given set $K_n$ of selected neighbors of $\bx_n$ (such as, e.g., the nearest neighbors, second nearest neighbors, etc., cf. Fig. \ref{fig:Kn}).
We first construct a smooth fitting function ${\hat \Phi}_{K_n} (\bx)$ of the values taken by $\hphi$ at the node set $K_n$. Next, we evaluate  ${\hat \Phi}_{K_n} (\bx) $ at the vertices ${\tilde \bx}_1,..., {\tilde \bx}_{\tilde N}$ of a second, structured triangulation ${\tilde \Pi}_h$, which is built up around ${\bx}_n$ (Fig. \ref{fig:Kn}).  We finally construct the following `regularized' polyhedral stress function

\bea
{\tilde \varphi}_h & = &  \sum_{n=1}^{\tilde N}  {\hat \Phi}_{K_n} ({\tilde \bx}_n) {\tilde g}_n
\label{th}
\eea 

}}

\noindent where ${\tilde N}$ is the number of nodes of ${\tilde \Pi}_h$, and ${\tilde g}_n$ denotes the piecewise linear basis function associated with ${\tilde \bx}_n \in {\tilde \Pi}_h$.
{\crev{Useful fitting models are offered by interpolation polynomials,
local maximum entropy shape function, Moving Least Squares (MLS)  meshfree approximations, and B-Splines, just to name a few examples (refer, e.g., to \cite{Cyr:2009} for a comparative study of such methods).
}}
Let us focus now on Eqns. (\ref{Pns}) and (\ref{Thn}).
The replacements of all the quantities relative to $\Pi_h$ with the analogous ones referred to ${\tilde \Pi}_h$ in such equations, leads us to (structured) `regularizations' ${\tilde {\mbs{P}}}_h$ and ${\tilde \bT}_h$ of the force network and stress field associated with the unstructured mesh $\Pi_h$, respectively.

\section{NUMERICAL RESULTS}
\label{numerics}

The present section provides a collection of numerical applications of the procedures described in the sections \ref{airy} and  \ref{stress}. 
We deal with the Flamant solution to the stress field of a half-plane loaded by a normal force, and tensegrity models of a cantilever beam and an elliptical dome.
In all the given examples, we analyze both structured and unstructured force networks describing the problem under examination, and study the properties of the associated stress fields.
Given a source triangulation $\Pi_s$, and a polyhedral function $\hphi$ associated with $\Pi_s$, we name \textit{smooth projection} of $\hphi$ over a target triangulation $\Pi_t$ the polyhedral function defined through: $(i$) the construction a smoothing of $\hphi$ through local quintic polynomials around each node of $\Pi_s$  \citep{Akima:1978}; $(ii)$ the sampling the fitting function ${\hat \Phi}$ at the vertices of $\Pi_t$. We assume that the fitting patch $K_n$ associated with such a projection coincides with the entire source mesh $\Pi_s$ (cf. Sect. \ref{stress}).

\subsection{Flamant problem}
\label{Flamant}

\begin{figure*}[htbp]
  \begin{center}
    \epsfig{file=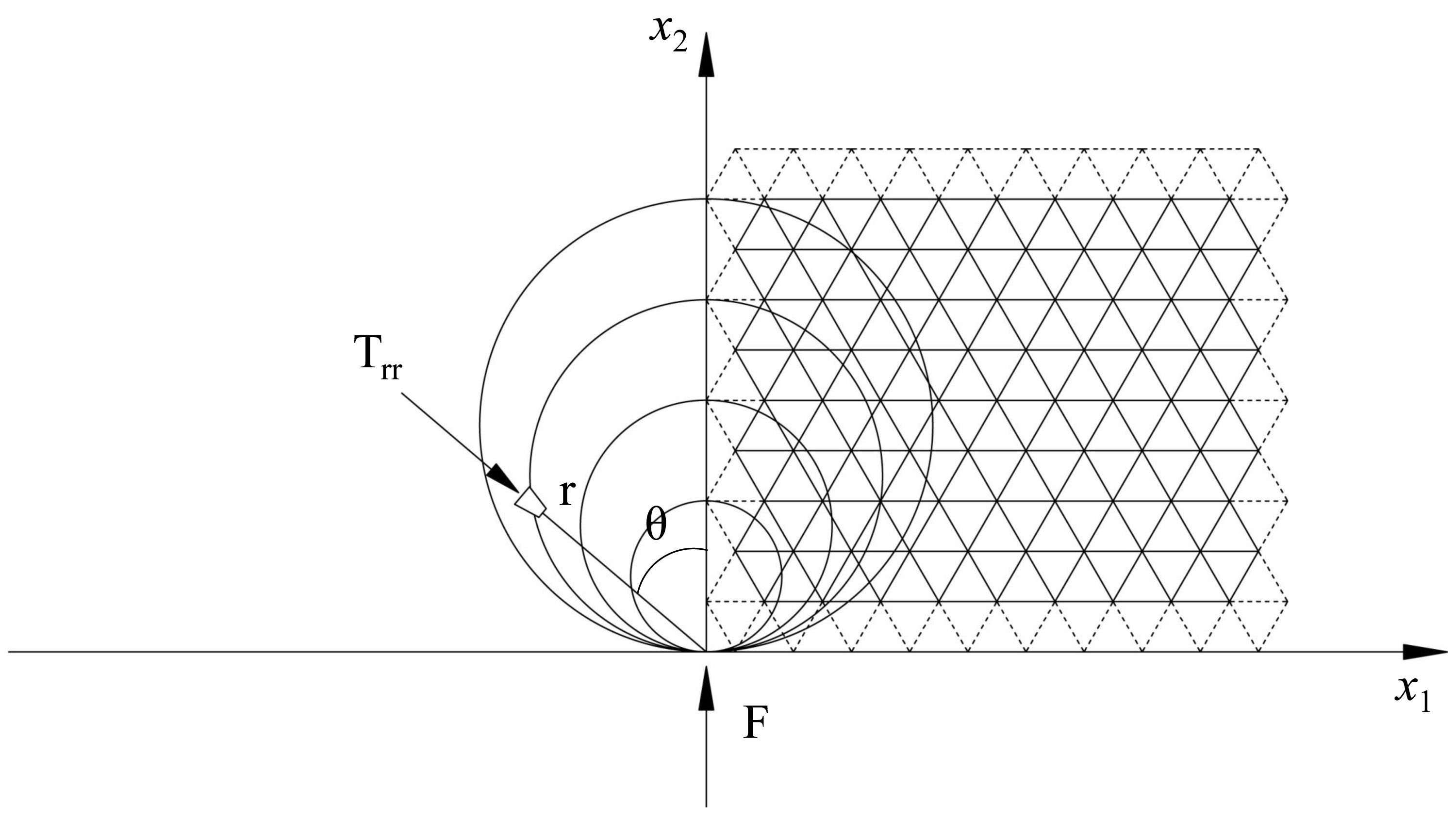,width=12cm} 
   \caption{Flamant solution for for the problem of a half plane loaded by a perpendicular point  load (left), and examined simulation region (right).}
    \label{fig:Flamant_problem}
  \end{center}
\end{figure*}

Let us study the convergence behavior of the regularized stress measure introduced in Sect. \ref{stress} by considering the well known Flamant solution for the problem of a half plane loaded by a perpendicular point  load. Such a problem has been analyzed in \cite{LSM2} through a lumped stress approach based on structured meshes. 
We examine the Flamant solution in terms of the Airy stress function, which reads

\bea
\varphi_0 & = & - \frac{F}{\pi} \ r \ \theta \ sin \theta
\label{Flamantphi}
\eea

\noindent where $r$ and $\theta$ are polar coordinates with origin at the point of application of the load $F$ (cf., e.g., \cite{LSM2}). The above stress function generates the following radial stress distribution in the loaded half-plane (Fig. \ref{fig:Flamant_problem}).

\bea
T_{rr}^{(0)} & = & - \frac{2 \ F \ cos \theta}{\pi \ r}
\label{FlamantTrr}
\eea

We consider approximations to $\varphi_0$ associated with four structured and unstructured triangulations of a $1.6 \times 1.4$ rectangular domain placed on one side of the loading axis (`simulation region', cf. Fig. \ref{fig:Flamant_problem}).
The analyzed structured triangulations  ${\tilde \Pi}^{(1)}, ..., {\tilde \Pi}^{(4)}$ are supported by hexagonal Bravais lattices, and show equilateral triangles with the following edge lengths: ${\tilde h}_1 = 0.20$ (mesh \# 1):  ${\tilde h}_2 = 0.10$ (mesh \# 2);
 ${\tilde h}_3 = 0.05$ (mesh \# 3); and  ${\tilde h}_4 = 0.025$ (mesh \# 4), respectively.
The unstructured triangulations $\Pi^{(1)},...,\Pi^{(4)}$ are instead obtained through random perturbations of the positions of the nodes of ${\tilde \Pi}^{(1)}, ..., {\tilde \Pi}^{(4)}$.

We first examine the projections ${\hat \varphi}^{(1)},..., {\hat \varphi}^{(4)}$ of the Flamant solution (\ref{Flamantphi}) over the unstructured meshes $\Pi^{(1)},...,\Pi^{(4)}$. Each of such stress functions generates an unstructured force network ${\hat {\mbs{P}}}^{(i)}$ (cf. Sect. \ref{airy}), and a piecewise constant approximation 
${\hat T}_{rr}^{(i)}$ to the Flamant stress field (Sect. \ref{stress}). 
Next, we construct a smooth projection ${\tilde \varphi}^{(i)}$ of the generic  ${\hat \varphi}^{(i)}$ over the structured triangulation ${\tilde \Pi}^{(i)}$ (\textit{unstructured to structured regularization}). We let ${\tilde {\mbs{P}}}^{(i)}$ and ${\tilde T}_{rr}^{(i)}$ respectively denote the force network and the discrete stress field associated with such a `regularized' stress function.

The accuracy of each examined approximation to the radial stress field (\ref{FlamantTrr}) is measured through the following Root Mean Square Deviation

\beq
\label{RMSD}
\begin{array}{lll}
\mbox{err}(T_{rr}^{(i)}) & = &  \sqrt{ \left( \sum^{N}_{n=1}{ \left( (T_{rr}^{(i)})_n - (T_{rr}^{(0)})_n\right)^2} \right) / N}
\end{array}
\eeq

\noindent where $N$ denotes the total number of nodes of the current mesh; $(T_{rr}^{(i)})_n$ denotes the value at node $n$ of ${T}_{rr}^{(i)}$; and $(T_{rr}^{0})_n$ denotes the value at the same node of the exact stress field (\ref{FlamantTrr}). In (\ref{RMSD}), we let ${T}_{rr}^{(i)}$ denote either ${\hat T}_{rr}^{(i)}$ (unstructured approximation to $T_{rr}^{(0)}$), or ${\tilde T}_{rr}^{(i)}$ (structured approximation to $T_{rr}^{(0)}$). 

\begin{figure*}[ht]
\unitlength1cm
\begin{picture}(14.0,15.75)
\put(5,15.4){${\hat {\mbs{P}}}^{(1)}$}
\put(11.5,15.4){${\tilde {\mbs{P}}}^{(1)}$}
\put(2.5,10.30){\epsfig{figure=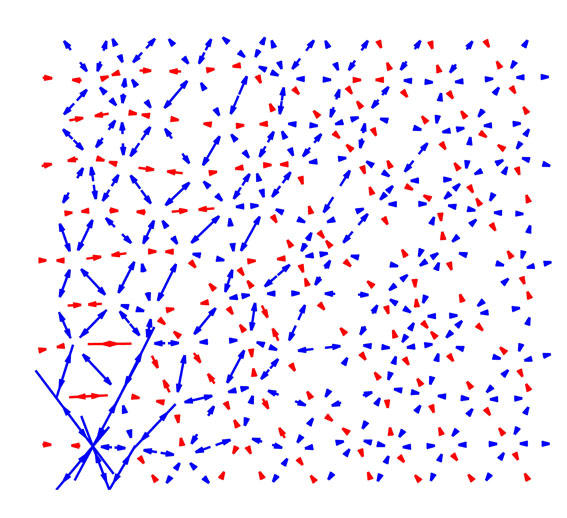,height=5cm}}
\put(8.75,10.30){\epsfig{figure=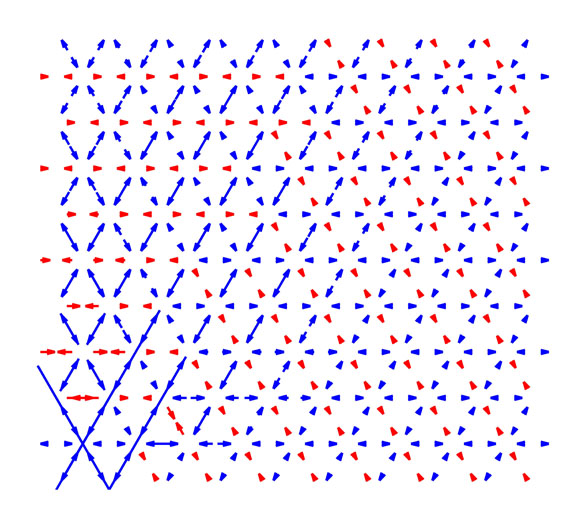,height=5cm}}
\put(5,9.95){${\hat {\mbs{P}}}^{(2)}$}
\put(11.5,9.95){${\tilde {\mbs{P}}}^{(2)}$}
\put(2.5,4.90){\epsfig{figure=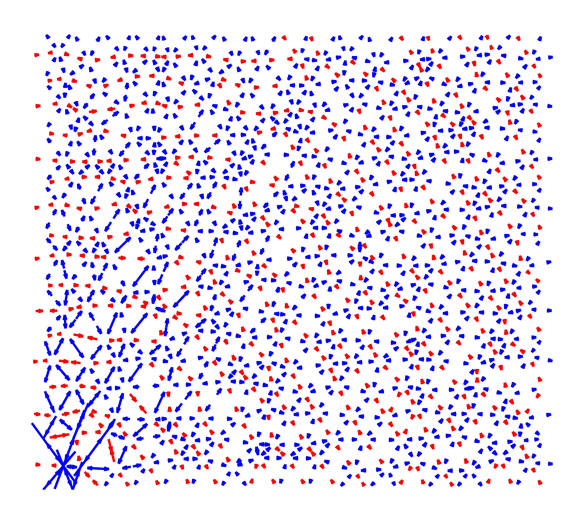,height=5cm}}
\put(8.755,4.90){\epsfig{figure=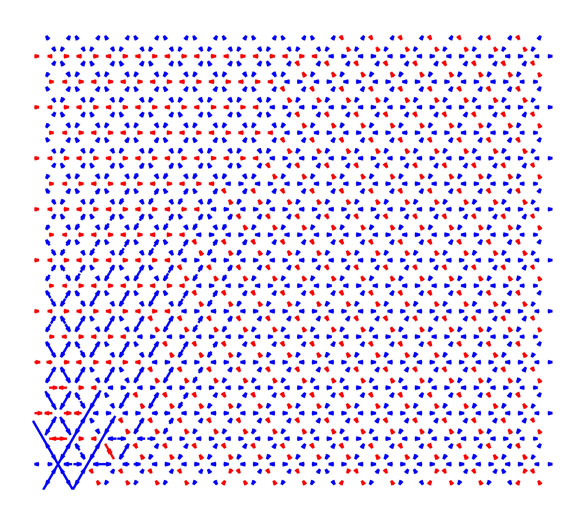,height=5cm}}
\put(5,4.55){${\hat {\mbs{P}}}^{(3)}$}
\put(11.5,4.55){${\tilde {\mbs{P}}}^{(3)}$}
\put(2.5,-0.5){\epsfig{figure=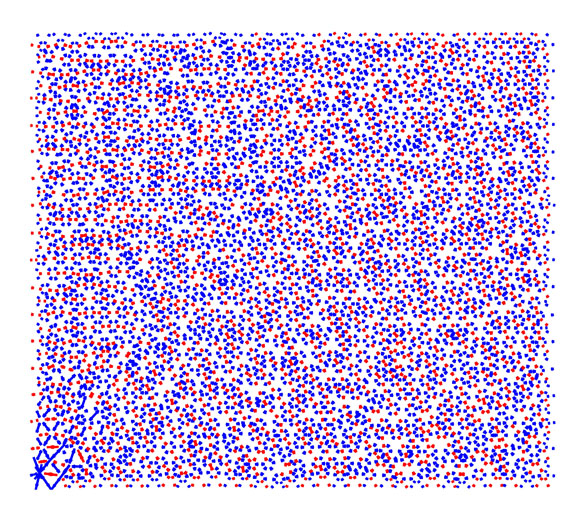,height=5cm}}
\put(8.75,-0.5){\epsfig{figure=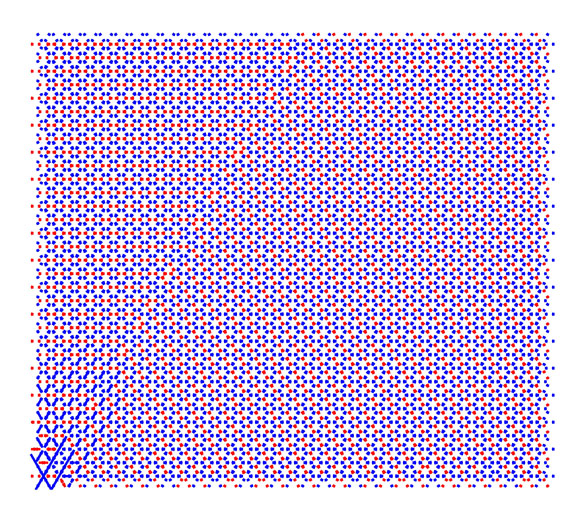,height=5cm}}
\end{picture}
\caption{(Color in the online version). Illustrations of selected unstructured (left) and structured (right) force networks approximating the Flamant problem in Fig. \ref{fig:Flamant_problem} (blue: compressive forces; red: tensile forces).}
\label{fig:Fmeshes}
\end{figure*}

\begin{figure*}[ht]
\unitlength1cm
\begin{picture}(14.0,6.0)
\put(1.0,0){\epsfig{figure=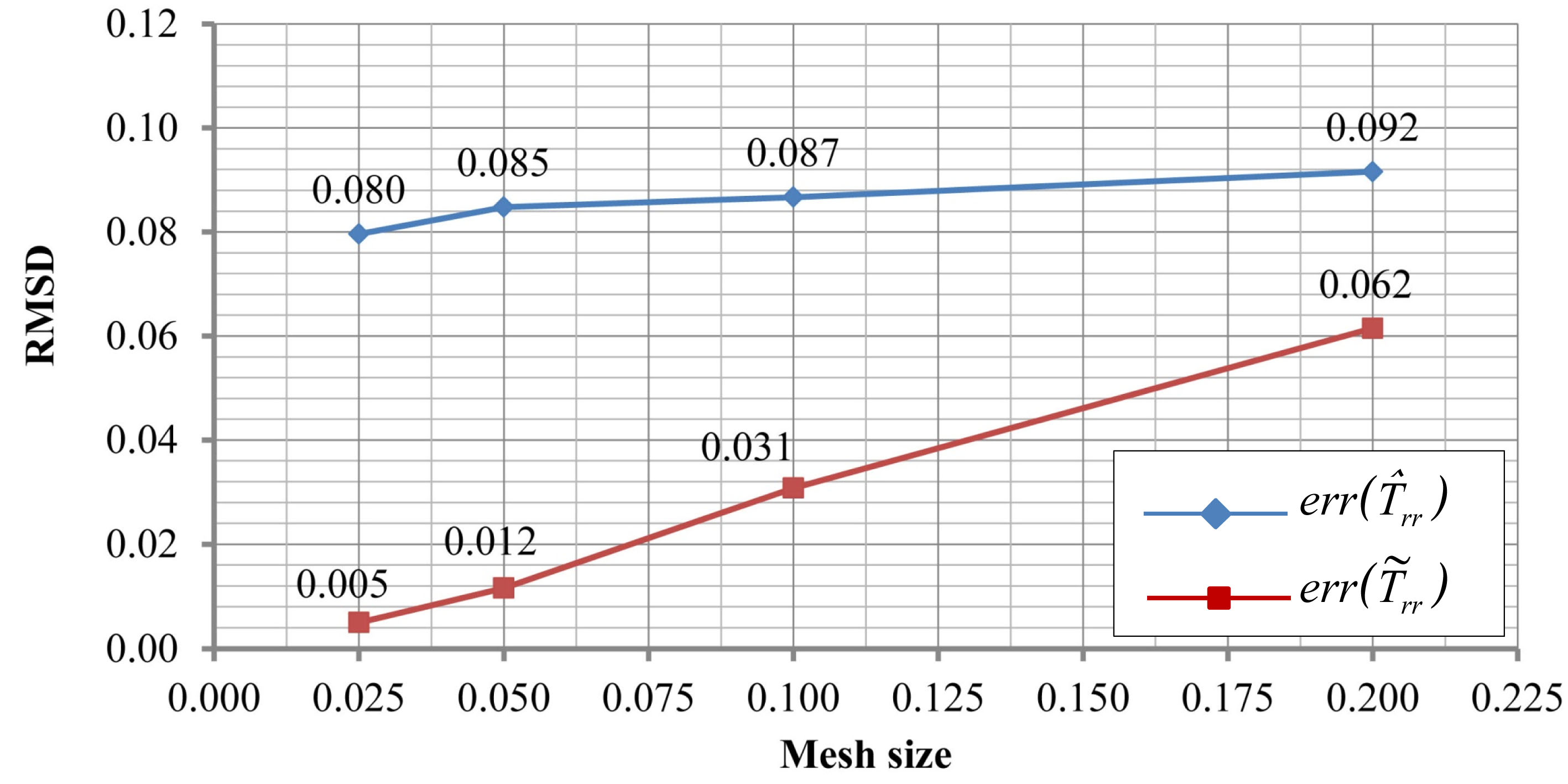,width=12cm}}
\end{picture}
\caption{(Color in the online version). Root Mean Square Deviations of the examined approximations to the radial stress $T_{rr}^{(0)}$ of the Flamant problem.}
\label{fig:Trr_err_Flamant}
\end{figure*}

\begin{figure*}[ht]
\unitlength1cm
\begin{picture}(14.0,11.0)
\put(1.5,6){\epsfig{figure=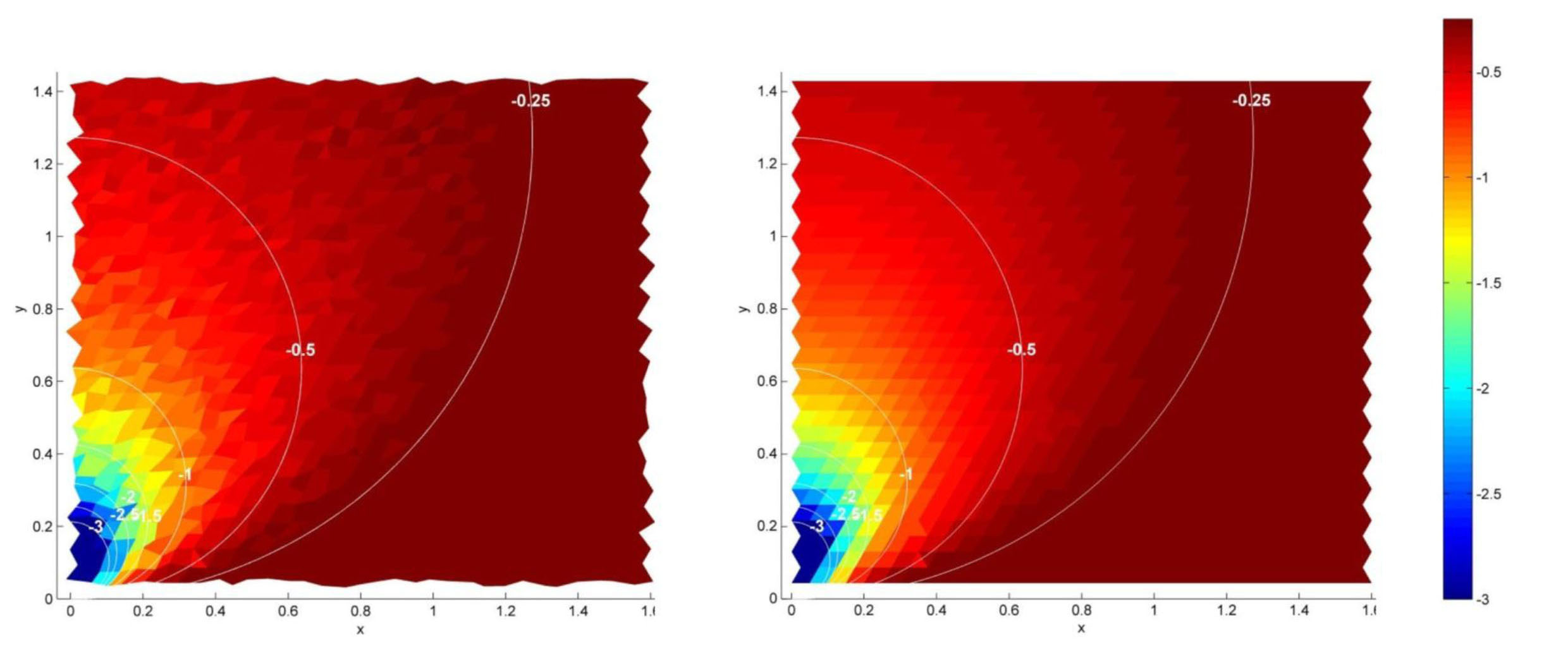,width=12cm}}
\put(4,11){${\hat T}_{rr}^{(3)}$}
\put(9.5,11){${\tilde T}_{rr}^{(3)}$}
\put(1.5,0.5){\epsfig{figure=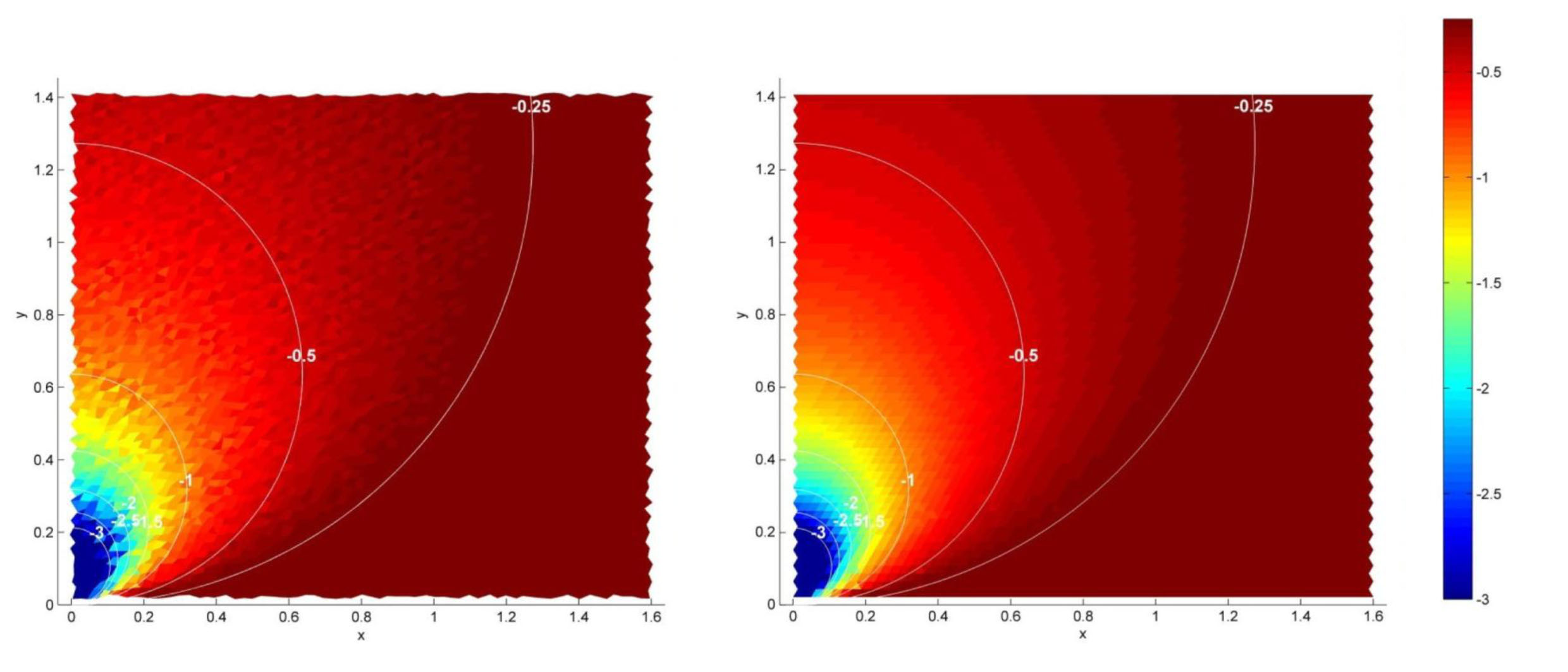,width=12cm}}
\put(4,5.25){${\hat T}_{rr}^{(4)}$}
\put(9.5,5.25){${\tilde T}_{rr}^{(4)}$}
\end{picture}
\caption{(Color in the online version). Density plots of the examined approximations to the radial stress $T_{rr}^{(0)}$ of the Flamant problem for different meshes and interpolation schemes.}
\label{Trr_plots_Flamant}
\end{figure*}

Fig. \ref{fig:Fmeshes}
graphically illustrates the force networks ${\hat {\mbs{P}}}^{(i)}$ and ${\tilde {\mbs{P}}}^{(i)}$ computed for some selected meshes, while
Fig. \ref{fig:Trr_err_Flamant} plots the approximation error (\ref{RMSD}) against the mesh size ${\tilde h}$, for each of the analyzed approximation schemes. Finally, Fig. \ref{Trr_plots_Flamant} depicts 3D density plots of ${\tilde T}_{rr}^{(i)}$ and ${\tilde T}_{rr}^{(i)}$ for meshes \#3 and \#4.
As the mesh size ${\tilde h}$ approaches zero, we observe from Fig. \ref{fig:Trr_err_Flamant} that the approximation errors of the unstructured approximations to $T_{rr}^{(0)}$ show rather low reduction rate, while those of the structured approximations instead feature slightly super-linear convergence to zero.
The results shown in Fig. \ref{Trr_plots_Flamant} confirm the higher degree of accuracy of the structured approximations ${\tilde T}_{rr}^{(i)}$, as compared to the unstructured approximations ${\hat T}_{rr}^{(i)}$. In this figure, we marked selected contour lines of the exact radial stress $T_{rr}^{(0)}$ by white circles  (cf. Fig. \ref{fig:Flamant_problem}).

\subsection{Cantilever truss}
\label{tensegrity}

\begin{figure*}[ht]
\unitlength1cm
\begin{picture}(14.0,13.0)
\put(2.0,9){\epsfig{figure=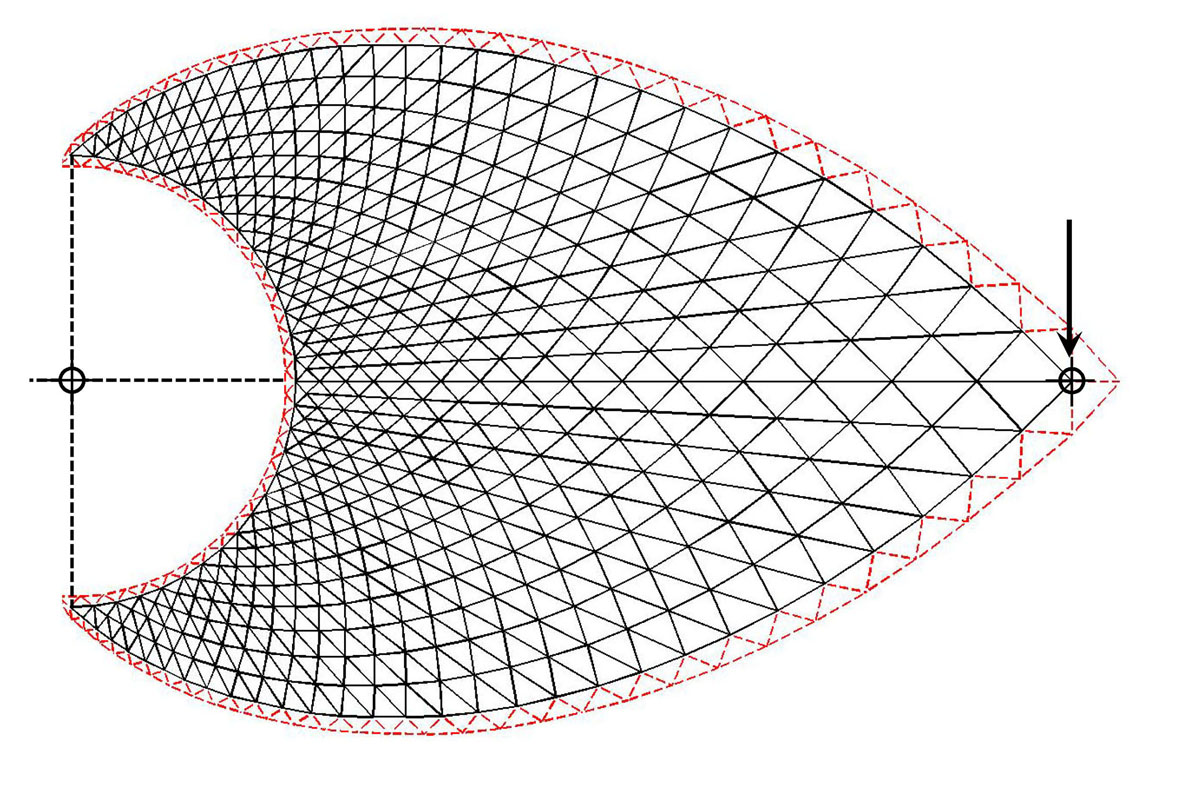,height=3.9cm}}
\put(5,13.0){${\tilde \Pi}$ }
\put(7.5,10.65){A}
\put(7.25,12.05){F}
\put(2.55,10.65){B}
\put(8.0,9){\epsfig{figure=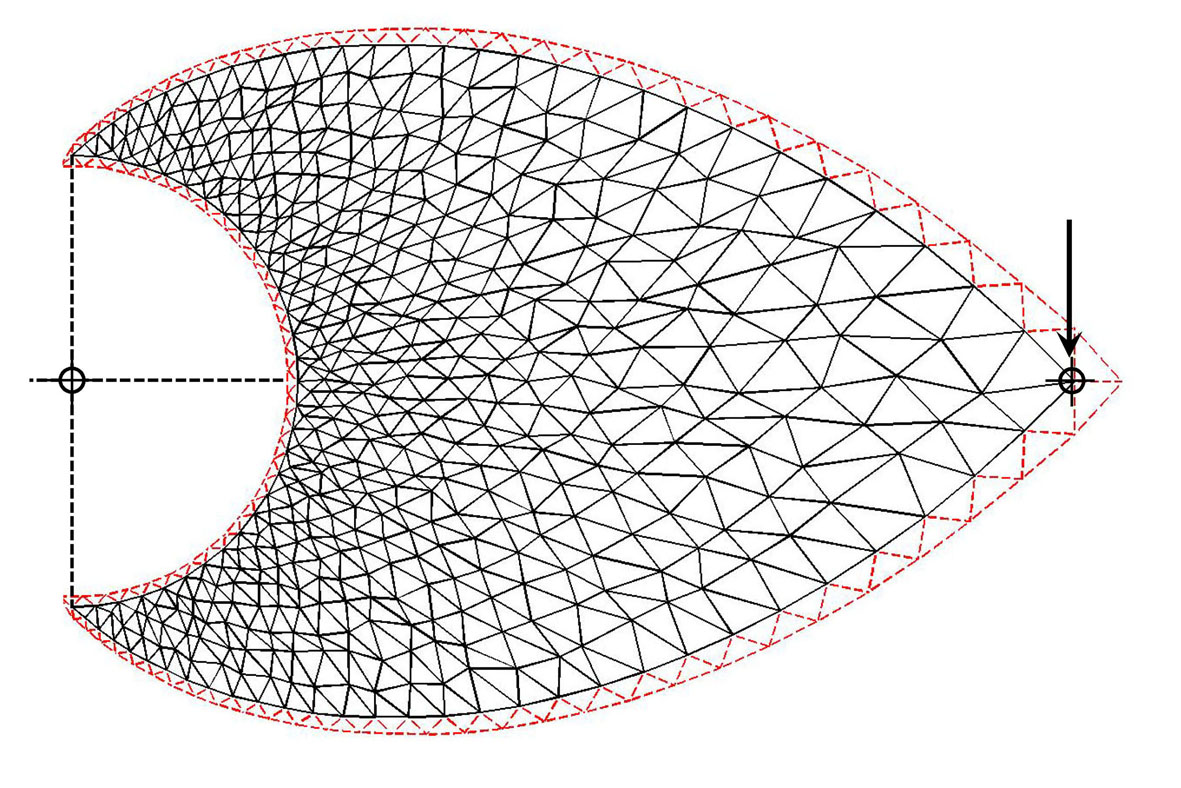,height=3.9cm}}
\put(11,13.0){${\Pi}$}
\put(13.5,10.65){A}
\put(13.25,12.05){F}
\put(8.55,10.65){B}
\put(2.5,4.0){\epsfig{figure=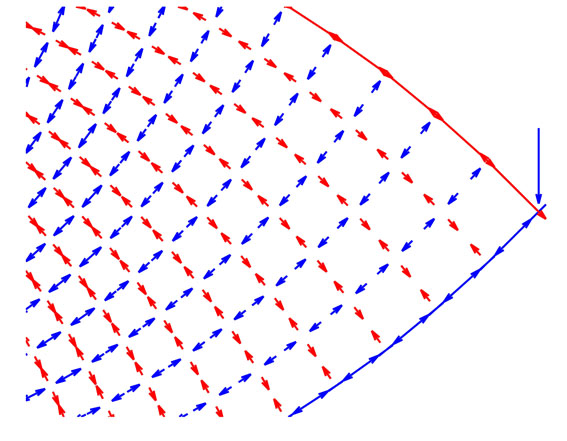,height=4cm}}
\put(5,8.25){${\tilde {\mbs{P}}}$}
\put(8.5,4.0){\epsfig{figure=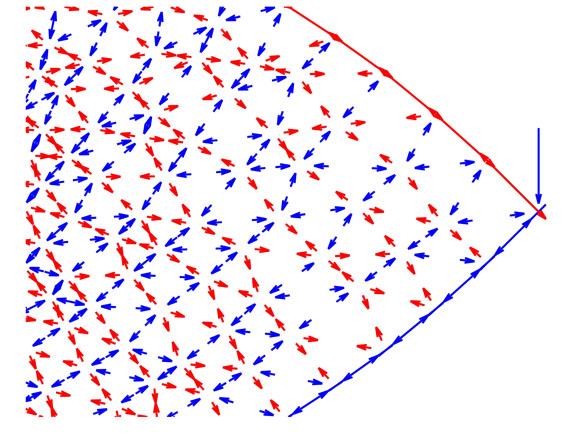,height=4cm}}
\put(11,8.25){${\hat {\mbs{P}}}$}
\put(1,0.0){\epsfig{figure=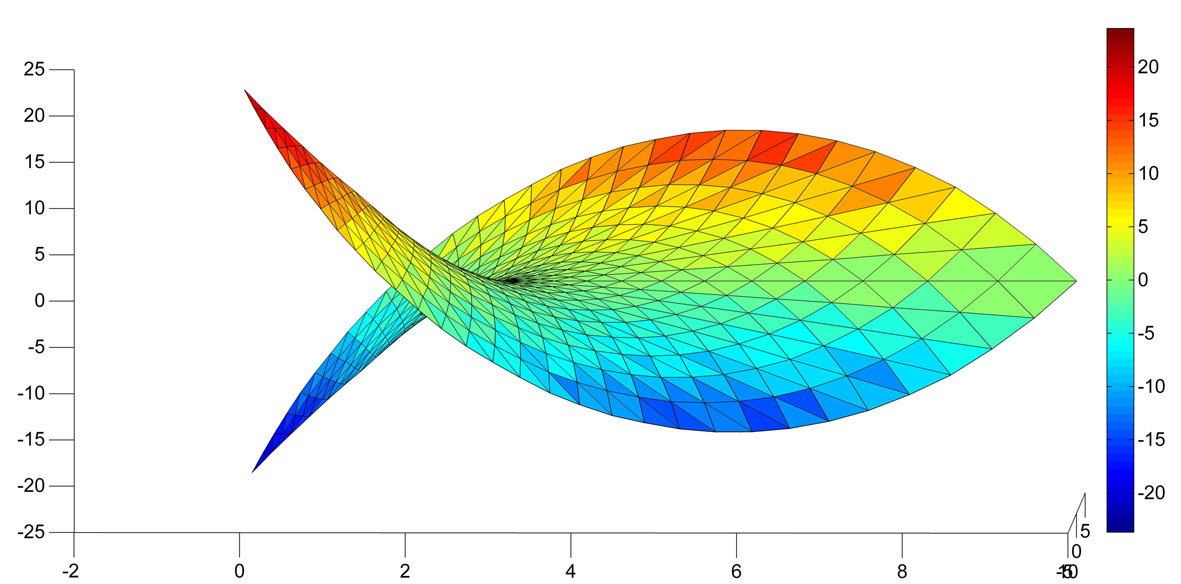,height=3.25cm}}
\put(5,3.5){${\tilde \varphi}$}
\put(8,0.0){\epsfig{figure=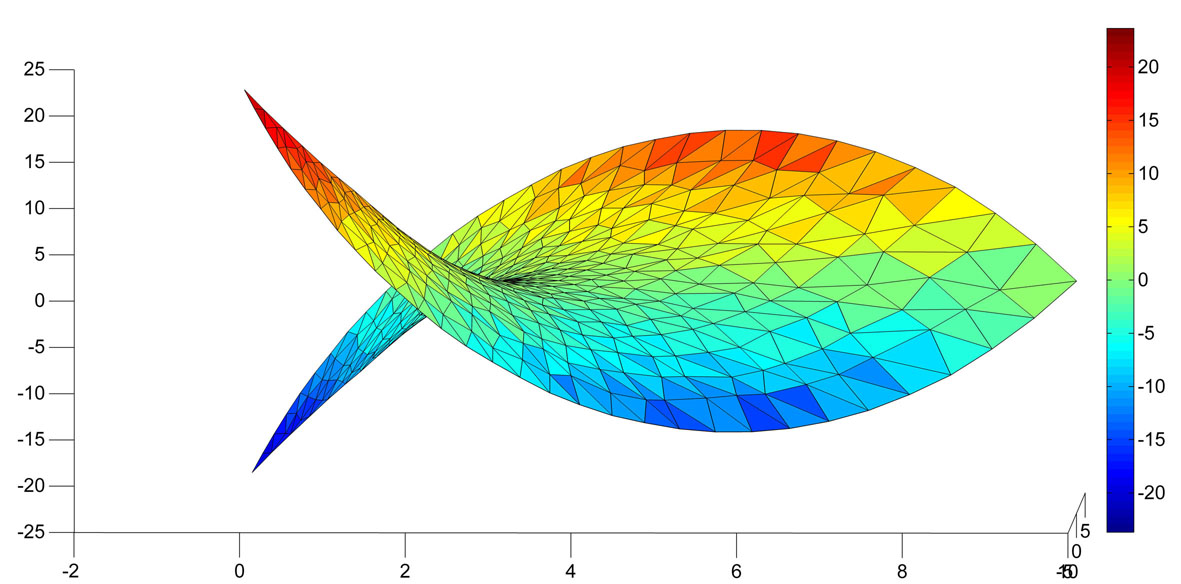,height=3.25cm}}
\put(11,3.5){${\hat \varphi}$}
\end{picture}
\caption{(Color in the online version). Michell truss example. Top: ordered (right) and unstructured (left) configurations. Center: details of the force networks near the tip (blue: compressive forces; red: tensile forces). Bottom: Airy stress functions associated with ordered (left) and unstructured (right) force networks.
}
\label{fig:Michell_trusses}
\end{figure*}

The current example is aimed to show how the procedures presented in Sects. \ref{airy} and \ref{stress} can be applied to determine the Airy stress function and the stress field associated with two different tensegrity models of a cantilever beam. 
We examine a truss structure ${\tilde \Pi}$ that has the same topology as the minimum volume frames analyzed in a famous study by A.G.M. Michell \cite{Michell} (see also \cite{Ske:2010}, Chap. 4).
Such a truss is composed of a system of orthogonal and equiangular spirals, which carries a force $F$ at a given point $A$, and is rigidly anchored in correspondence with a small circle centered at the origin $B$ of the spirals (refer to Fig. \ref{fig:Michell_trusses}, and \cite{Michell, Ske:2010, Prager:Michell}). We assume that the length of the $AB$ segment is $10$; the opening angle of the truss is $\pi$; the radius of the anchoring circle is $2$; and it results $F=10$ (in abstract units). 
We complete the Michell truss with the insertion of diagonal edges connecting the two orders of spirals, obtaining an enriched truss model supported by a triangulation with 589 nodes and 1578 physical edges (cf. Fig. \ref{fig:Michell_trusses}).
We also consider a perturbed configuration ${\Pi}$ of the Michell truss, which is obtained by randomly moving the inner nodes of the regular configuration (Fig. \ref{fig:Michell_trusses}). 

\begin{figure*}[ht]
\unitlength1cm
\begin{picture}(14.0,10.5)
\put(7.0,5.5){\epsfig{figure=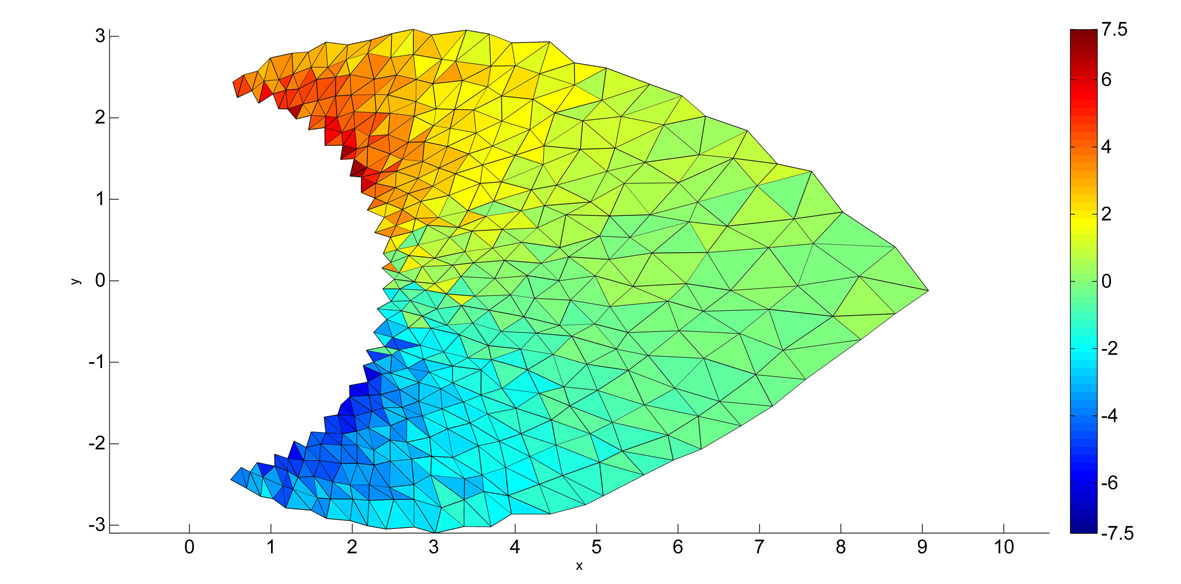,height=4cm}}
\put(0.0,5.5){\epsfig{figure=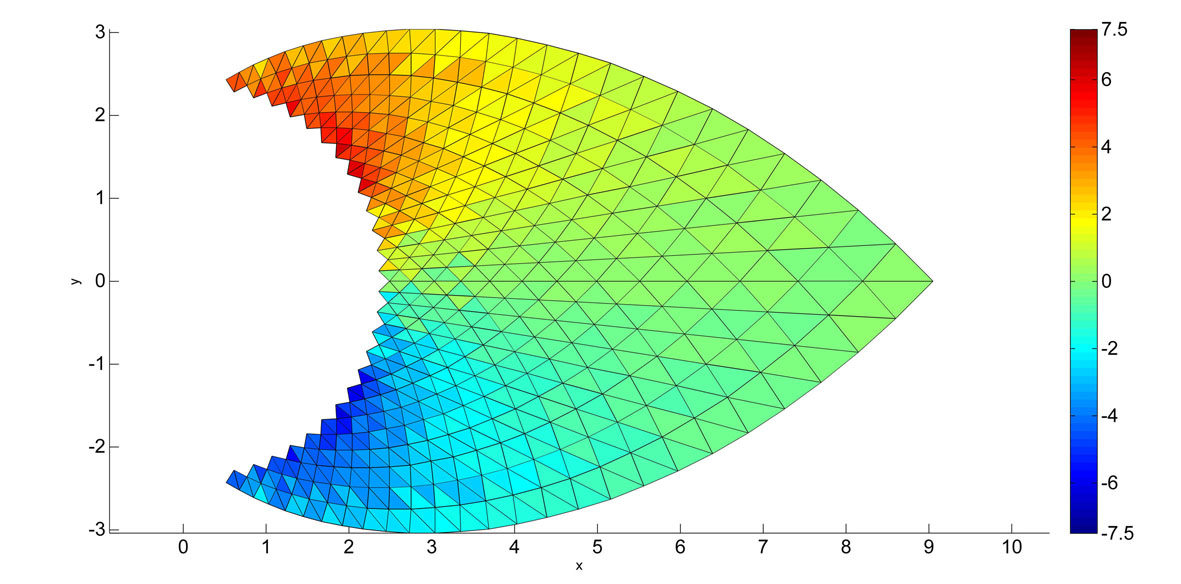,height=4cm}}
\put(7.0,0.5){\epsfig{figure=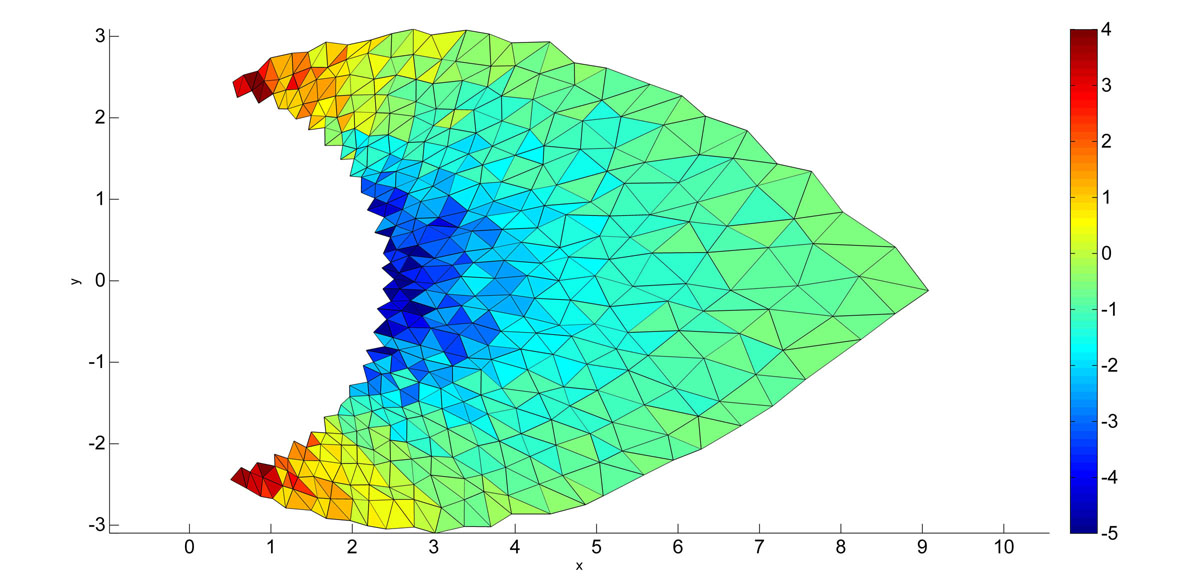,height=4cm}}
\put(0.0,0.5){\epsfig{figure=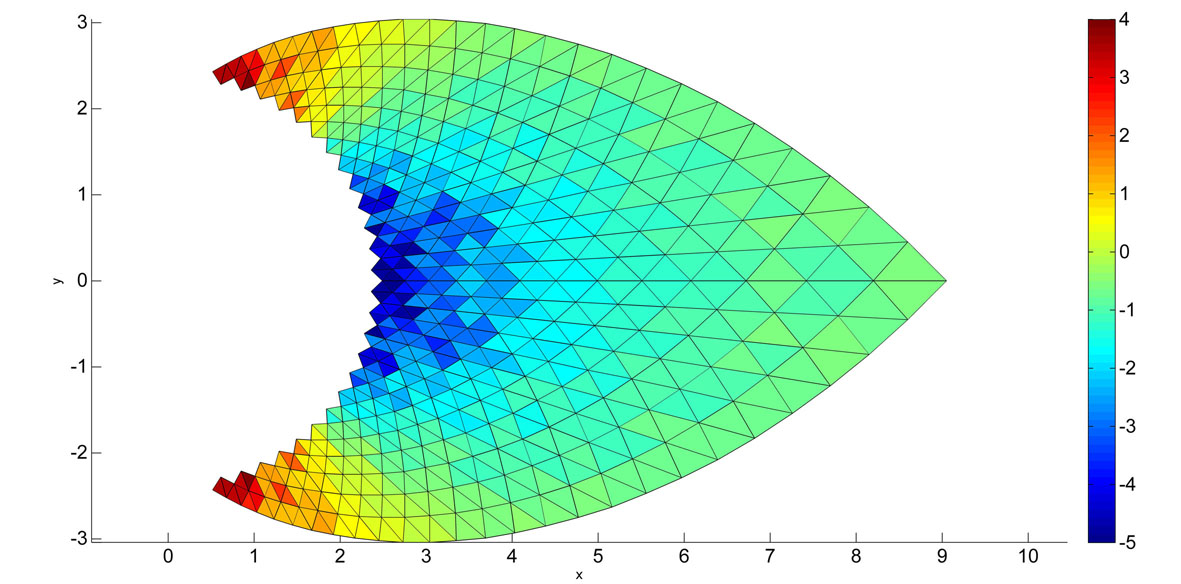,height=4cm}}
\put(4,9.7){${\tilde T}_{11}$}
\put(11,9.7){${\hat T}_{11}$}
\put(4,4.7){${\tilde T}_{12}$}
\put(11,4.7){${\hat T}_{12}$}
\end{picture}
\caption{(Color in the online version). Density plots of different approximations to the stress components $T_{11}$ (top:longitudinal normal stresses) and $T_{12}$ (bottom:tangential stresses) associated with the Michell truss.}
\label{fig:Michell_T11}
\end{figure*}
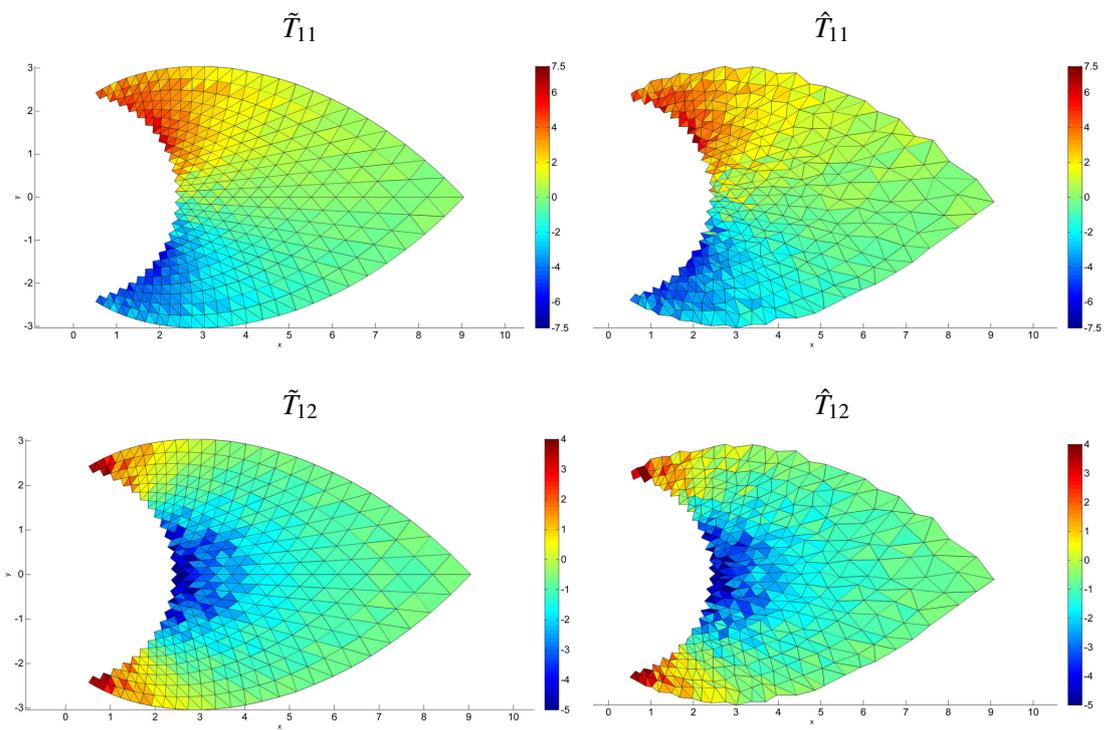

We initially follow Michell's approach to the equilibrium problem of ${\tilde \Pi}$, by computing the axial forces in the spiral members through the nodal equilibrium equations of the structure (refer to \cite{Ske:2010}, Chap. 4), and setting the forces in the remaining edges to zero (`Michell truss'). Next, we associate an Airy stress function ${\tilde \varphi}$ to such a force network ${\tilde {\mbs{P}}}$, through Eqn. (\ref{Pinv})  of Sect. \ref{forces-airy} (cf. Fig. \ref{fig:Michell_trusses}). 
On proceeding in reverse ordered with respect to the previous example, we then construct a smooth projection 
${\hat \varphi}$ 
of ${\tilde \varphi}$ over the  perturbed configuration ${\Pi}$, and let ${\hat {\mbs{P}}}$ denote the associated force network (Fig. \ref{fig:Michell_trusses}). 
Let us focus our attention on the Cartesian components $T_{11}$ and $T_{12}$ of the stress fields associated with ${\tilde {\mbs{P}}}$ and ${\hat {\mbs{P}}}$ ($x_1$ denoting the longitudinal axis).
The results in Fig. \ref{fig:Michell_T11}  highlight that the `structured stress' ${\tilde {\mbs{T}}}$ (associated with ${\tilde {\mbs{P}}}$) smoothly describes the stress field associated with the background domain of the Michell truss, while the `unstructured stress' ${\hat \bT}$ (associated with ${\hat {\mbs{P}}}$), on the contrary, provides a fuzzy description of such a stress field. 

\begin{figure*}[ht]
\unitlength1cm
\begin{picture}(14.0,15.0)
\put(3.5,10.0){\epsfig{figure=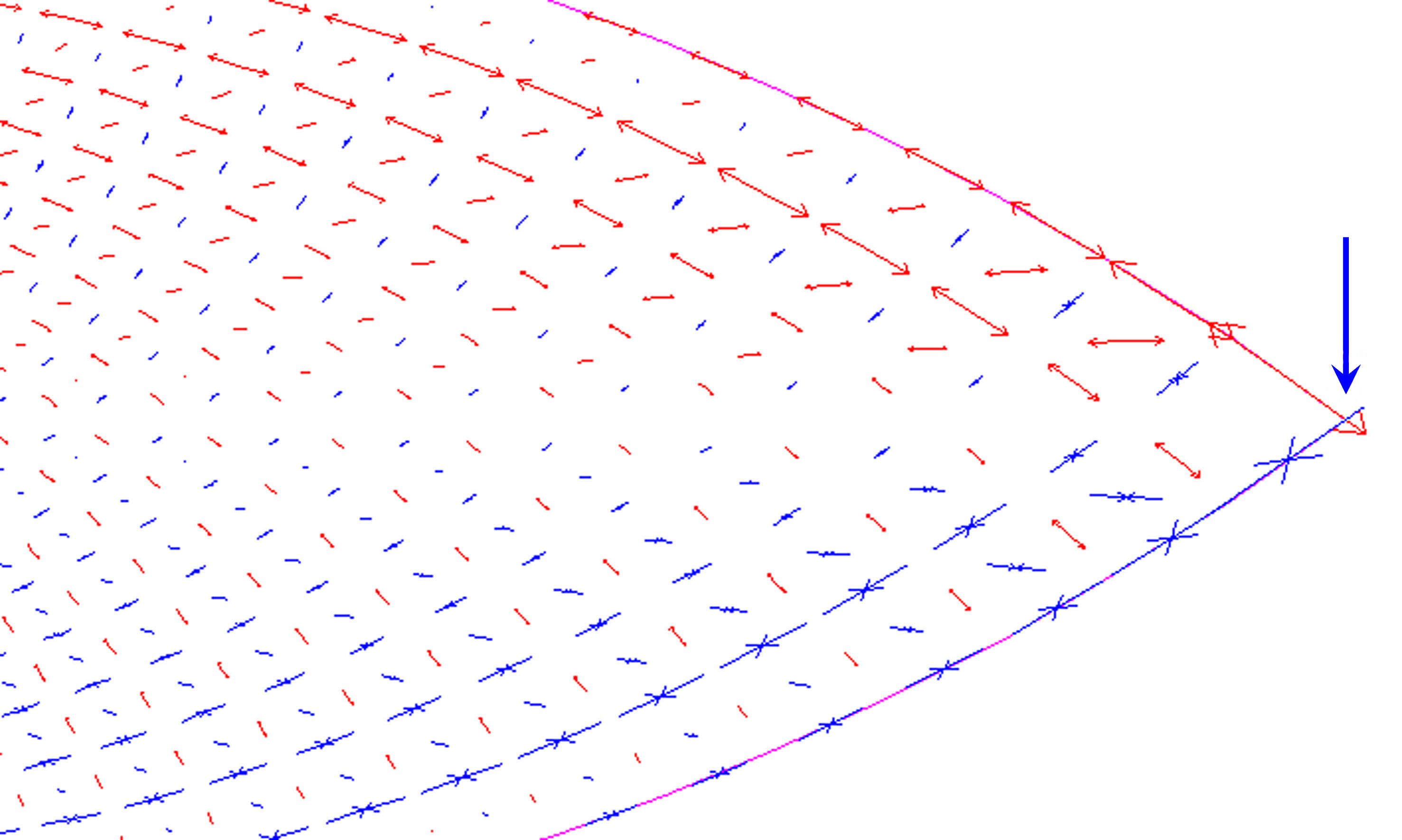,height=3.25cm,width=4cm}}
\put(5,13.75){${\tilde {\mbs{P}}}$ }
\put(9.5,10.0){\epsfig{figure=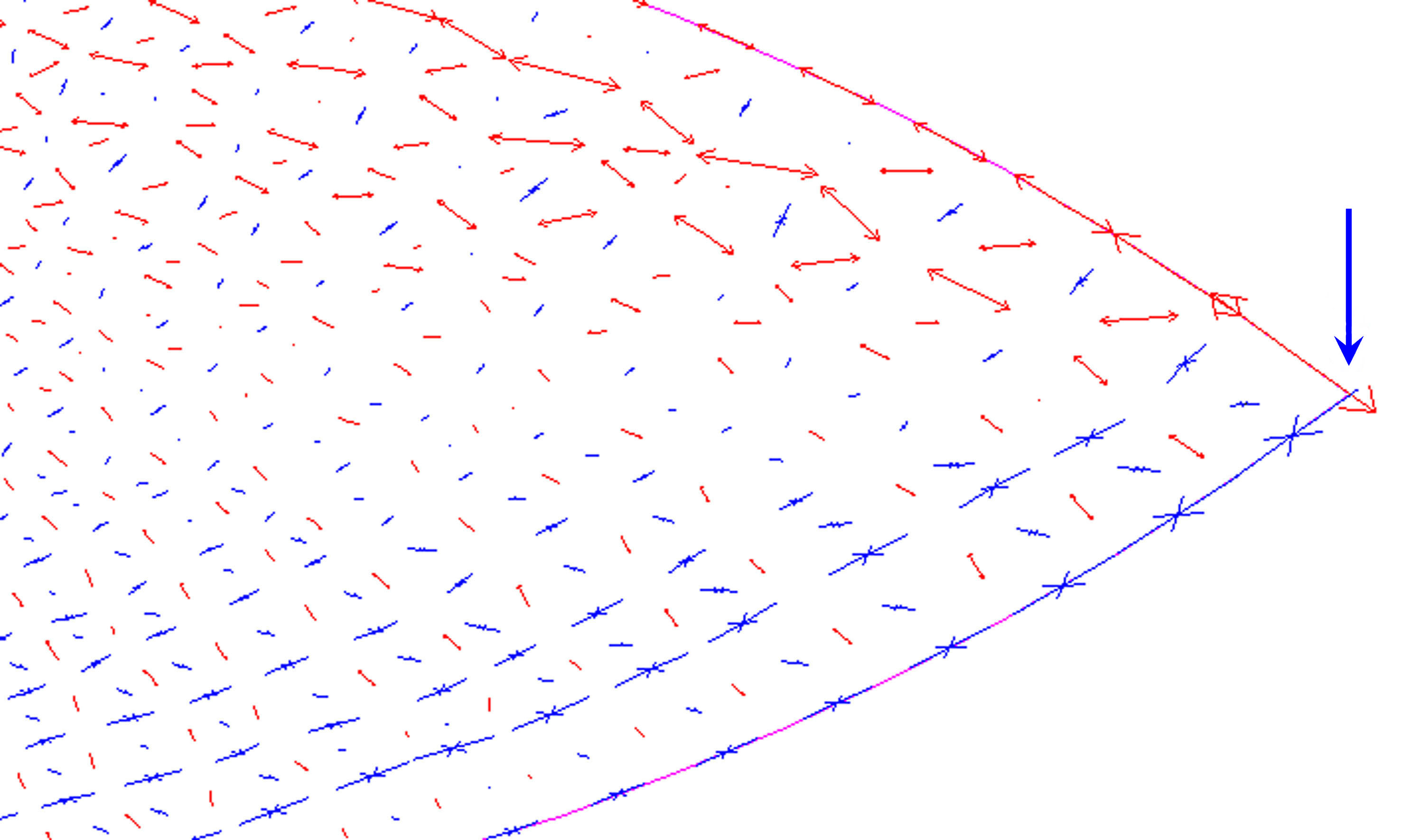,height=3.25cm,width=4cm}}
\put(11,13.75){${\hat {\mbs{P}}}$}
\put(2.5,5.0){\epsfig{figure=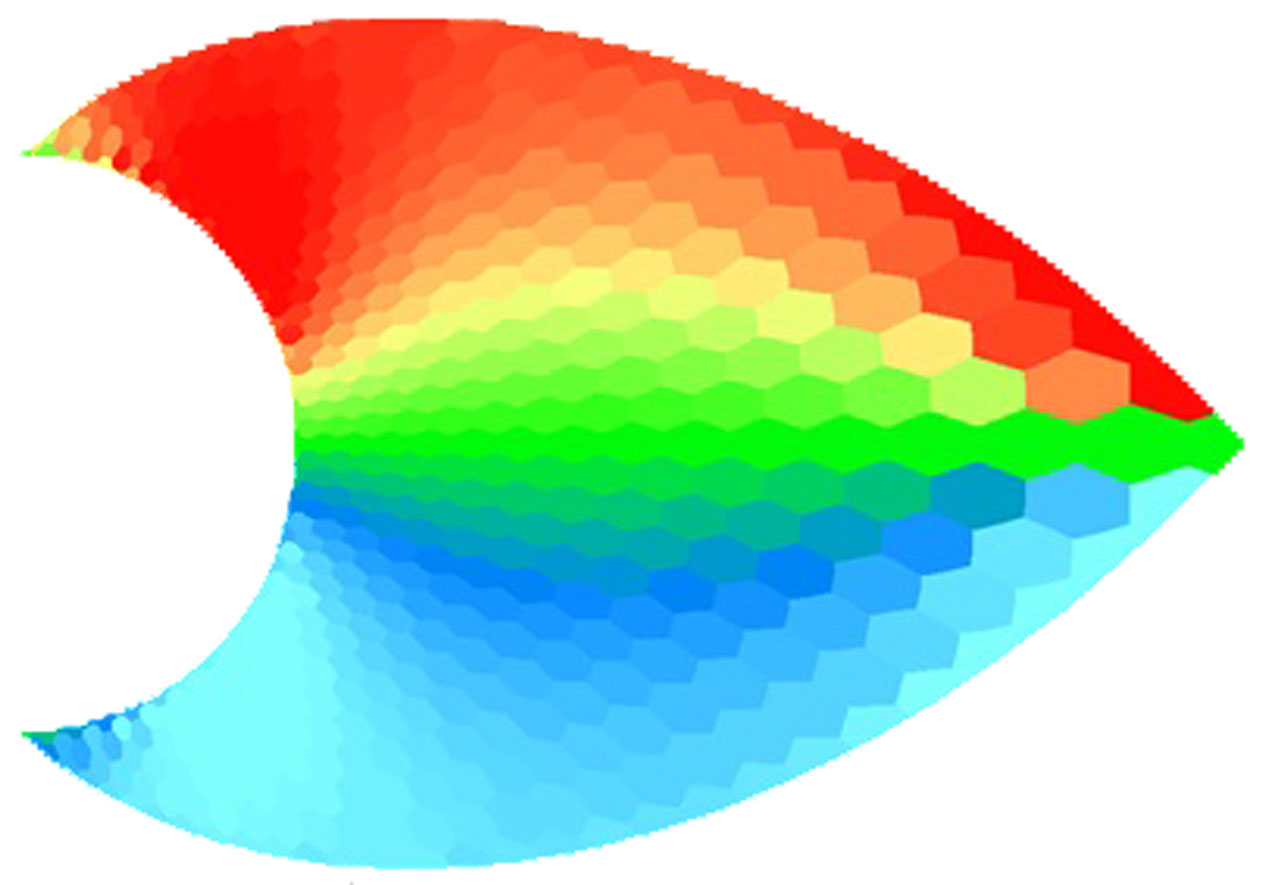,height=4cm}}
\put(5,9.25){${\tilde T}_{11}$}
\put(8.5,5.0){\epsfig{figure=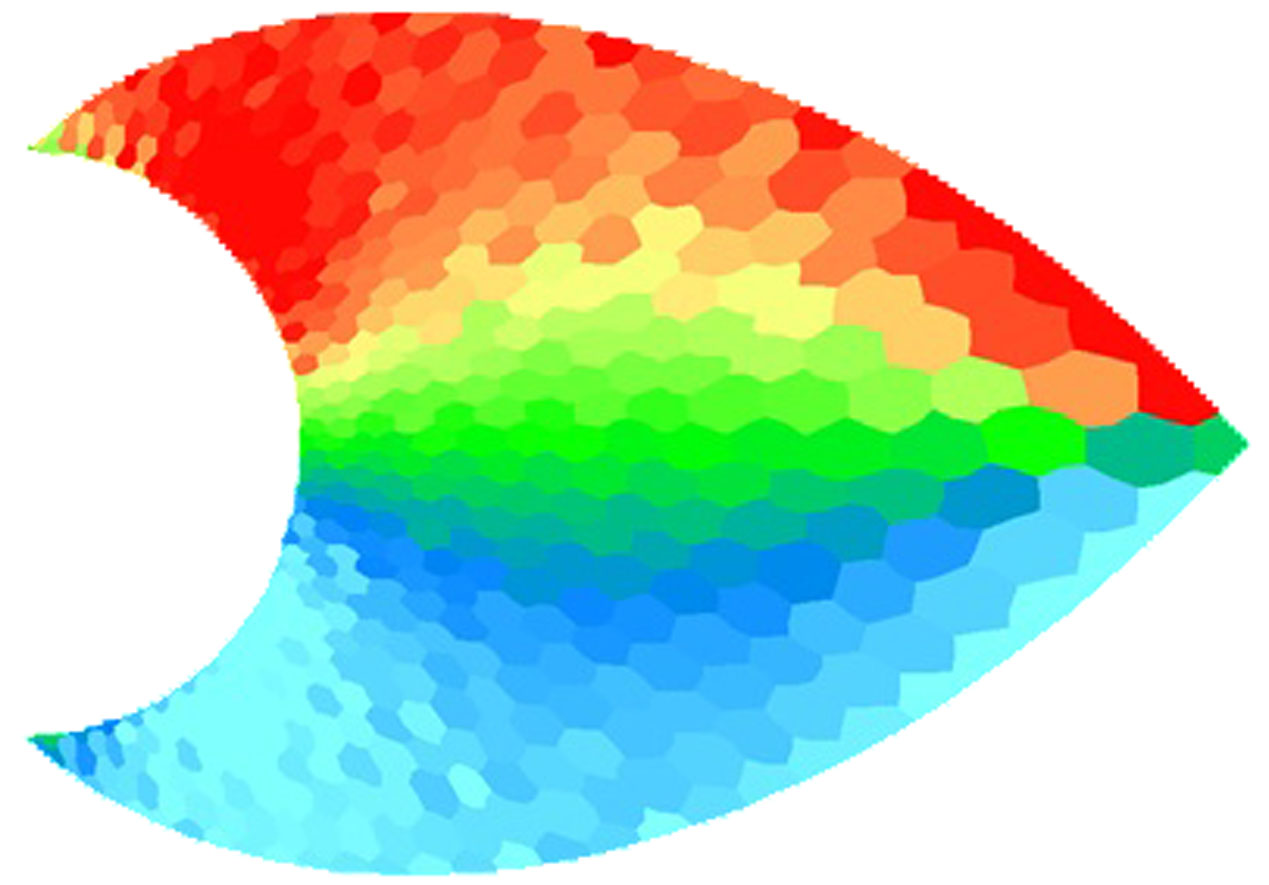,height=4cm}}
\put(14.5,5.0){\epsfig{figure=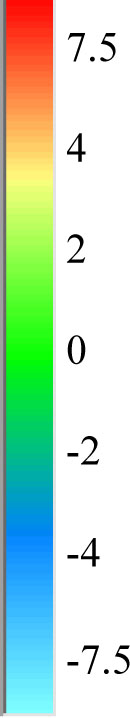,height=4cm}}
\put(11,9.25){${\hat T}_{11}$}
\put(2.5,0.0){\epsfig{figure=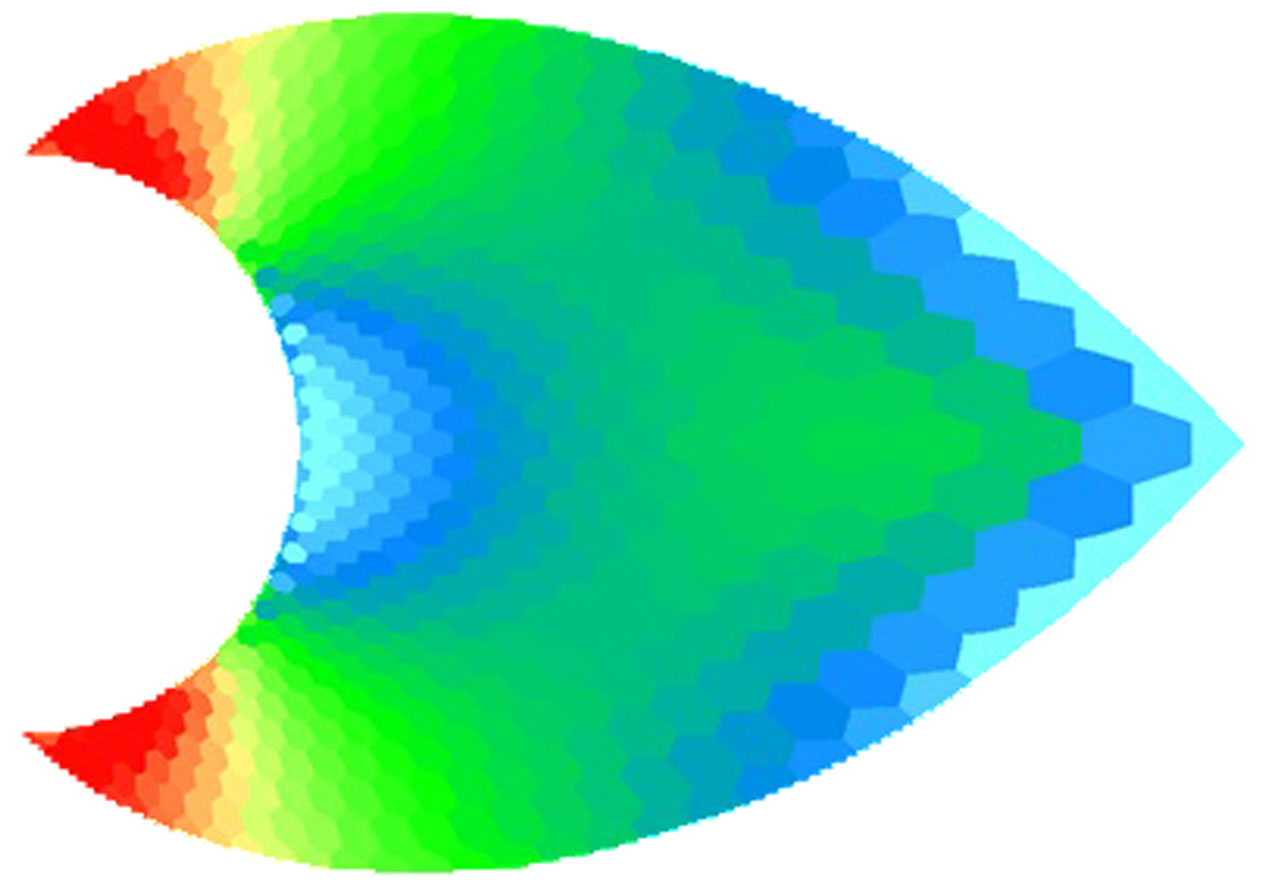,height=4cm}}
\put(5,4.25){${\tilde T}_{12}$}
\put(8.5,0.0){\epsfig{figure=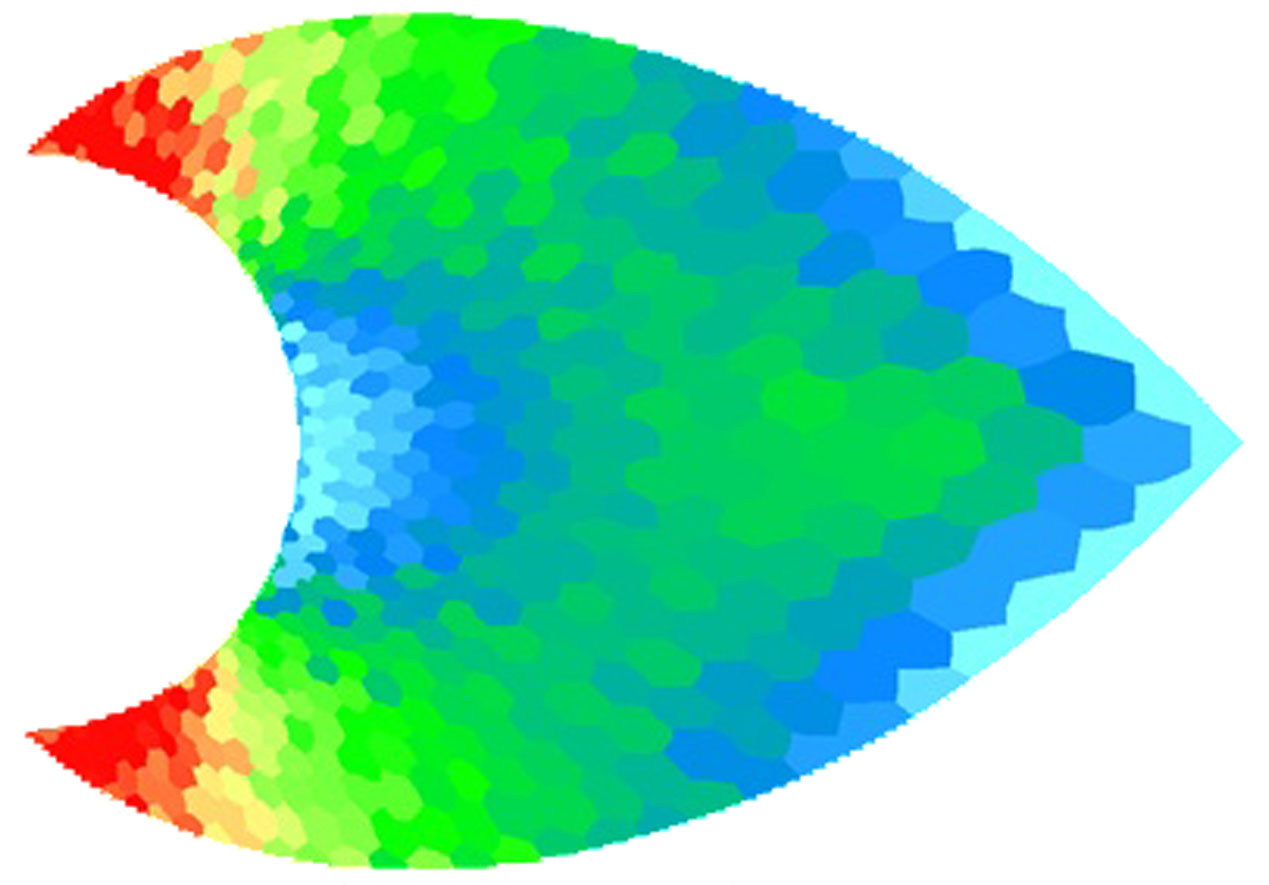,height=4cm}}
\put(14.5,0){\epsfig{figure=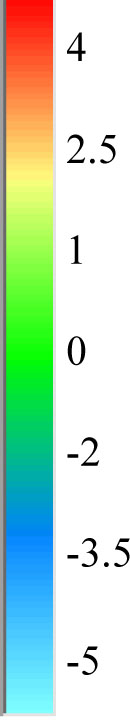,height=4cm}}
\put(11,4.25){${\hat T}_{12}$}
\end{picture}\caption{(Color in the online version). LSM truss example. Top: ordered (left) and unstructured (right) force networks (blue: compressive forces; red: tensile forces). Center and bottom: ordered and unstructured approximations to the stress field of the background domain.
}
\label{Michell_LSM}
\end{figure*}

A different approach to the truss ${\tilde \Pi}$ is obtained by looking at
the 2D elastic problem of the background domain $\Omega$ (here supposed to be homogeneous), under the given boundary conditions.
We now interpret ${\tilde \Pi}$ as a lumped stress model of $\Omega$, i.e., a
non-conventional elastic truss having the strain energy computed per nodes (i.e., per dual elements) and not per elements (`LSM truss', cf. \cite{LSM2}). 
Accordingly, we determine  the forces in its members by solving the elastic problem presented in Sect. 5 of \cite{LSM2}.
As in the previous case, we also consider the smooth projection of the Airy function associated with the regular truss ${\tilde \Pi}$ over the perturbed configuration ${\Pi}$. 
We show in Fig. \ref{Michell_LSM} the force networks and the stress fields corresponding to the LSM trusses ${\tilde \Pi}$ (Fig. \ref{Michell_LSM}, left), and ${\Pi}$ (Fig. \ref{Michell_LSM}, right). 
By comparing the results in Figs. \ref{fig:Michell_trusses} and \ref{fig:Michell_T11}  with those in Fig. \ref{Michell_LSM}, we realize that the LSM truss ${\tilde \Pi}$ shows non-zero forces in the non-spiral members, differently from the Michell truss (Fig. \ref{fig:Michell_trusses}, left). 
The results in Figs. \ref{fig:Michell_T11} and \ref{Michell_LSM} 
point out that averaging techniques based on unstructured force networks do not generally produce smooth descriptions of the Cauchy stress field, as we already observed in Sect. \ref{stress}. 

\bigskip

\subsection{Elliptical dome}
\label{elliptical dome}

Our last example is concerned with a tensegrity model of a membrane shaped as an elliptic paraboloid.
The membrane equilibrium problem of such a structure can be approached by Pucher's theory (refer, e.g., to \cite{Mansfield:1964, Baratta:2010, LSM4}), which introduces a stress function $\varphi_0$ to generate projected membrane stresses $T_{\alpha \beta}$ (Pucher stresses) onto the horizontal  plane (membrane `platform'). We examine an elliptical dome described by the following Monge chart

\bea
x_3 & = &  \left(1 - \frac{x_1^2}{a^2} -  \frac{x_2^2}{b^2} \right) h
\label{zdome}
\eea

\noindent where $x_1$ and $x_2$ are Cartesian coordinates in the plane of the platform; $x_3$ is the coordinate orthogonal to such a plane; $h$ is the maximum rise; and $a$ and $b$ are the semi-axes of the elliptic basis. On considering the stress function defined by

\bea
\varphi_0 & = &  - \frac{q a^2 b^2}{4 h} \left( sin \left( \frac{x_1 \pi}{a} \right)  \  sin \left( \frac{x_2 \pi}{b} \right)  \right)
\label{phidome}
\eea

\noindent we generate the following Pucher stresses over the membrane platform

\bea
T_{11}^{(0)} & = &  \frac{q a^2 \pi^2}{4 h} \left( sin \left( \frac{x_1 \pi}{a} \right)  \  sin \left( \frac{x_2 \pi}{b} \right)  \right)  \nn \\
T_{22}^{(0)} & = &   \frac{q b^2 \pi^2}{4 h} \left( sin \left( \frac{x_1 \pi}{a} \right)  \  sin \left( \frac{x_2 \pi}{b} \right)  \right)  \\
T_{12}^{(0)} & = &  \frac{q a b \pi^2}{4 h} \left( cos \left( \frac{x_1 \pi}{a} \right)  \  cos \left( \frac{x_2 \pi}{b} \right)  \right) \nn
\label{Tdome}
\eea

\begin{figure*}[ht]
\unitlength1cm
\begin{picture}(14.0,13.0)
\put(0.5,9){\epsfig{figure=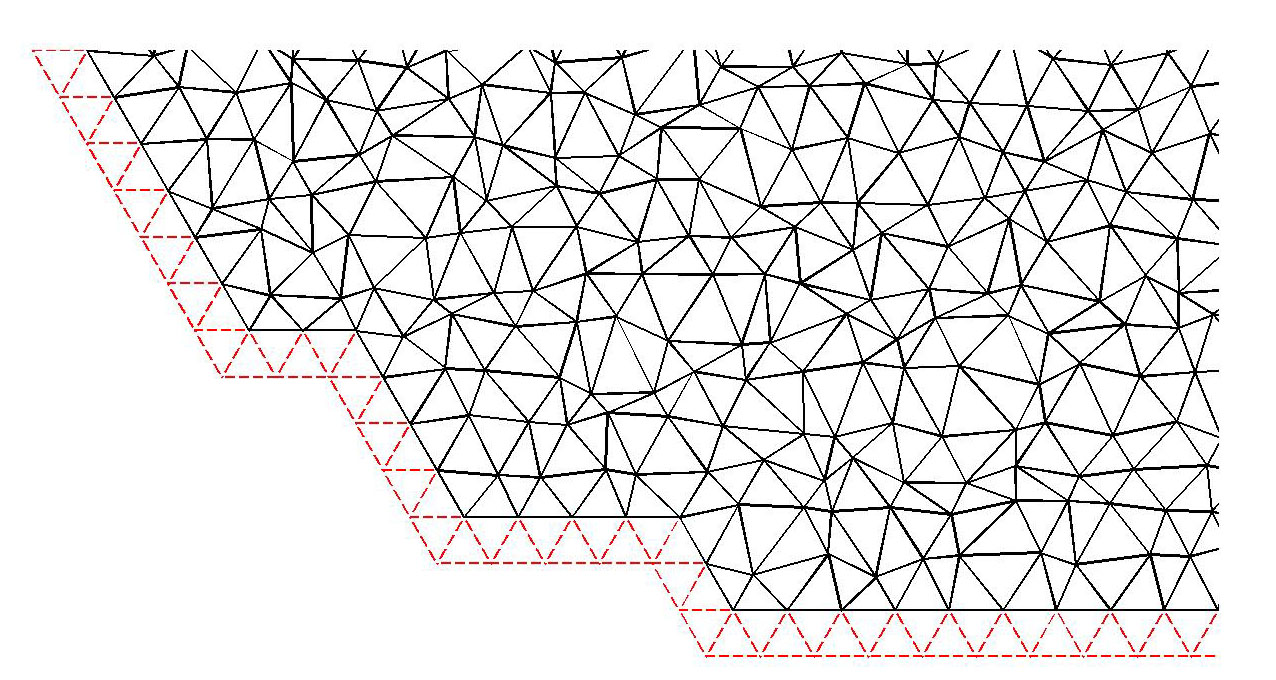,height=4cm}}
\put(4.0,12.85){${\Pi}$ }
\put(8.5,9){\epsfig{figure=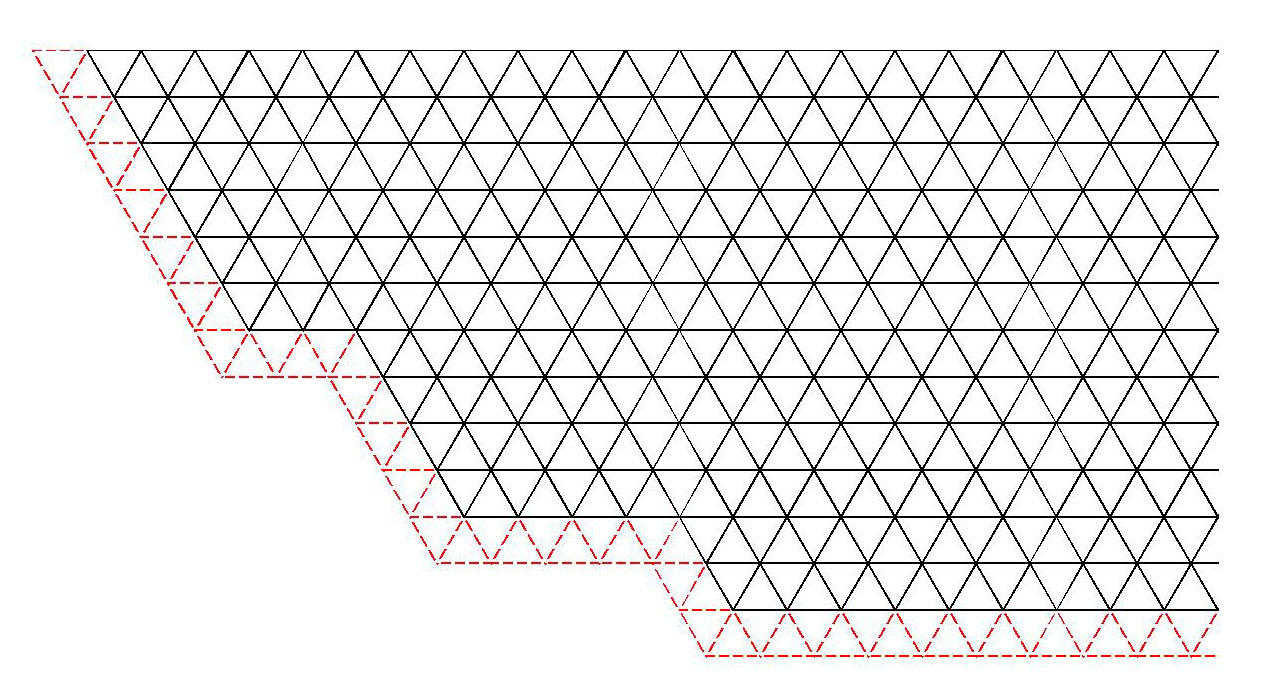,height=4cm}}
\put(11.75,12.85){${\tilde \Pi}$ }
\put(0.5,4.5){\epsfig{figure=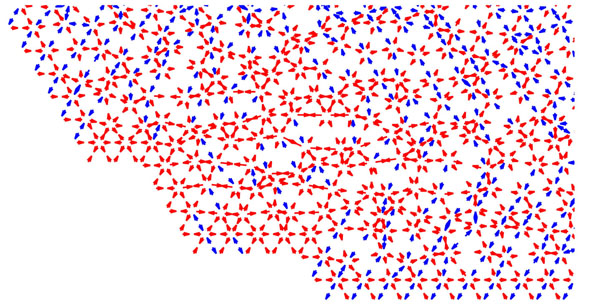,height=3.75cm}}
\put(4,8.35){${\hat {\mbs{P}}}$ }
\put(8.5,4.5){\epsfig{figure=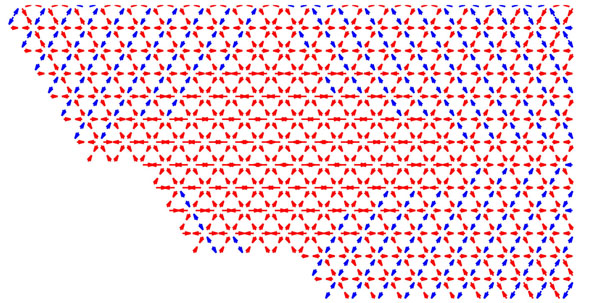,height=3.75cm}}
\put(11.75,8.35){ ${\tilde {\mbs{P}}}$ }
\put(1,0){\epsfig{figure=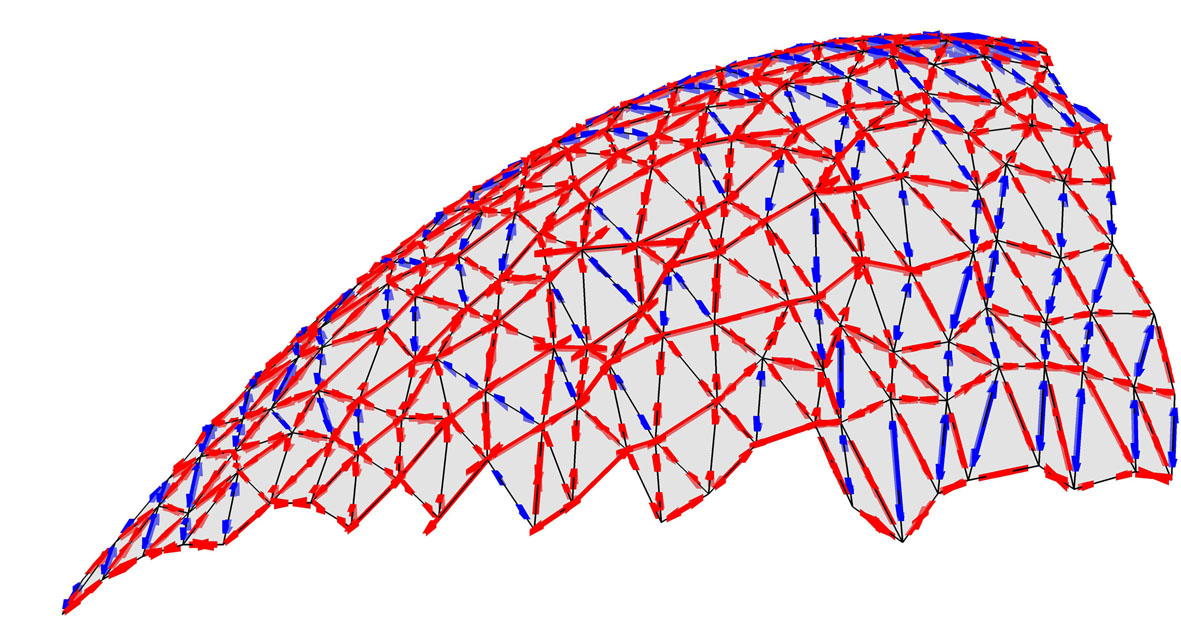,height=3.5cm}}
\put(4,3.8){${\hat {\mbs{N}}}$ }
\put(9,0){\epsfig{figure=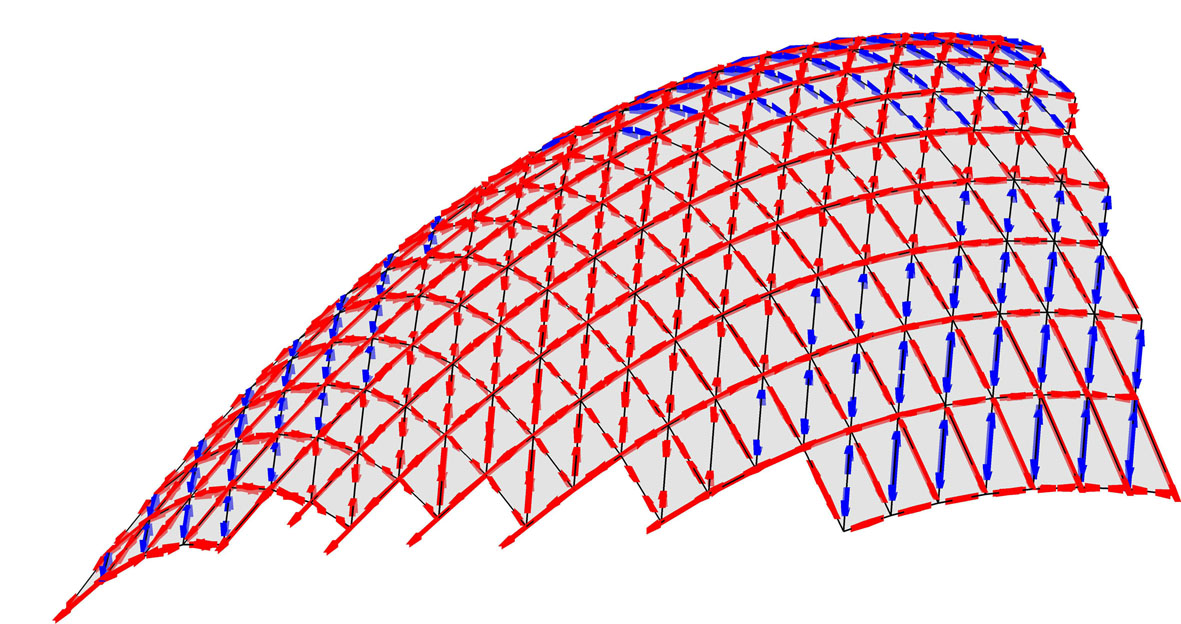,height=3.5cm}}
\put(11.75,3.8){${\tilde {\mbs{N}}}$ }
\end{picture}
\caption{(Color in the online version). Adopted meshes (top); 2D force networks (center); and 3D force networks (bottom) for a quarter of unstructured (left) and structured (right) tensegrity models of an elliptic dome.
 }
\label{fig:Emeshes}
\end{figure*}

\begin{figure*}[ht]
\unitlength1cm
\begin{picture}(14.0,13.5)
\put(0.5,0.0){\epsfig{figure=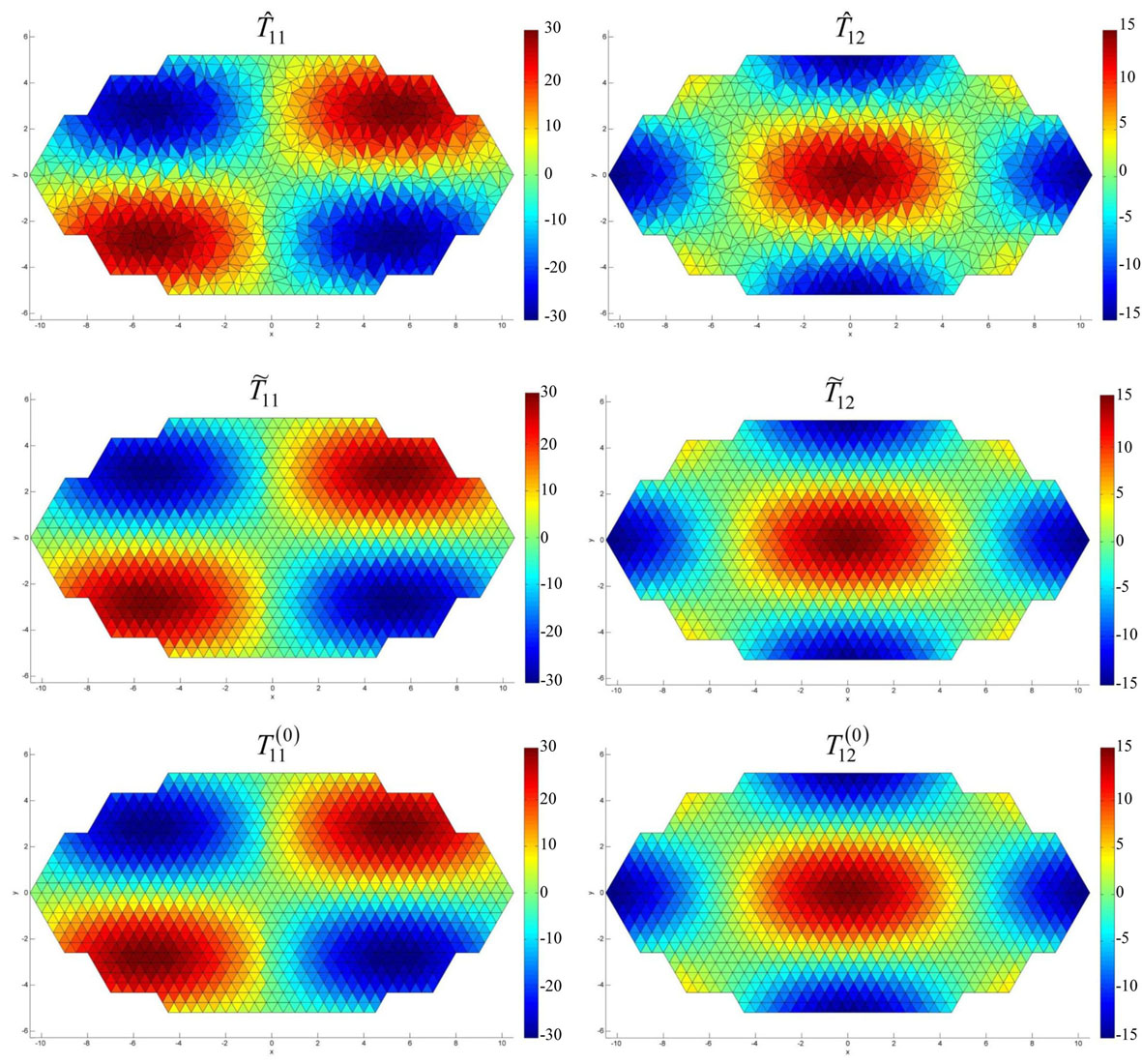,height=14cm}}
\end{picture}
\caption{Density plots of the examined approximations to the Pucher stresses $T_{11}^{(0)}$ and $T_{12}^{(0)}$ for the elliptic dome example.
}
\label{fig:Estresses}
\end{figure*}

\noindent Let us assume $q=1$, $a=11.26$, $b=5.63$, $h=10$ (in abstract units). 
As in the previous example, we study a structured and an unstructured tensegrity models of the problem under examination.
The structured model ${\tilde \Pi}$ is supported by a hexagonal Bravais lattice featuring 953 nodes and 2628 physical edges, while the unstructured model ${\Pi}$ is obtained through random perturbations of the positions of the nodes of ${\tilde \Pi}$. In both cases, we approximate 
the elliptic basis of the dome by a polygon with 22 edges (cf. Figs. \ref{fig:Emeshes} and \ref{fig:Estresses}).
In the present example, we first project the stress function (\ref{phidome}) over the unstructured triangulation $\Pi$, 
and denote the corresponding polyhedral function by ${\hat \varphi}$.
Next, we construct a smooth projection ${\tilde \varphi}$ of  ${\hat \varphi}$ over the structured triangulation ${\tilde \Pi}$ (\textit{unstructured to structured regularization}). The force networks ${\hat {\mbs{P}}}$ and ${\tilde {\mbs{P}}}$, which are respectively associated with ${\hat \varphi}$ and ${\tilde \varphi}$, define unstructured and a structured tensegrity models of the platform.
We can easily transform such force systems into 3D force networks ${\hat {\mbs{N}}}$ and ${\tilde {\mbs{N}}}$, by lifting the vertices of the meshes ${\Pi}$ and ${\tilde \Pi}$ at the  height of the surface (\ref{phidome}), respectively (Fig. \ref{fig:Emeshes}). 
Let ${\hat T}_{\alpha \beta}$ and ${\tilde T}_{\alpha \beta}$ denote the unstructured and structured approximations to the Pucher stresses (\ref{Tdome}), which correspond to the force networks ${\hat {\mbs{P}}}$ and ${\tilde {\mbs{P}}}$, respectively (Sect. \ref{stress}).
The density plots in Fig. \ref{fig:Estresses} show clear evidence for a close match between the structured stresses ${\tilde T}_{11}$, ${\tilde T}_{12}$ and the exact Pucher stresses ${T}_{11}^{(0)}$, ${T}_{12}^{(0)}$, and the `fuzzy' aspect of the unstructured stresses ${\hat T}_{11}$, ${\hat T}_{12}$.

\section{CONCLUDING REMARKS}
\label{conclusions}

We have formulated and discussed the relationship between polyhedral Airy stress function and internally self-equilibrated frameworks of simply connected domains in two dimensions, by generalizing classical results of plane elasticity \cite{Green:2002, Gurtin}.
We have also formulated a two-mesh technique for the definition of the Cauchy stress associated with unstructured force networks, which handles arbitrary triangulations of simply-connected domains, and makes use of smooth projection operators.
The results of Sect. \ref{numerics} highlight that the smooth projection of an unstructured stress function over a structured triangulation is able to generate a convergent discrete notion of the Caucy stress in the continuum limit.
Such a stress measure can be usefully employed to smoothly predict the stress field associated with tensegrity models of flat and curved membranes \cite{Dwy:1999, LSM1, LSM2, DaviniParoni:2003, Kil:2005, Blo:2007, Blo:2009, Micheletti:2008, LSM4, LSM5, Blo:2011, Angelillo:2012, Desbrun:2013, DeGoes:2013}, and to formulate concurrent discrete-continuum approaches based on the lumped stress method \cite{LSM1, LSM3, LSM4, LSM5, LSM2}. Due to its ability in generating  unstructured and structured force networks over a given design domain, the proposed regularization technique can also be used in association with structural optimization procedures and form-finding methods \citep{LSM4, LSM5, Lin:2010, Ohmori:2011, Wang:2010, Baldassini:2010}.

Several aspects of the present work pave the way to relevant further investigations and generalizations that we address to future work. First, the inclusion of body forces calls for specific attention, since network structures are usually loaded by nonzero forces at all nodes. 
Such a generalization of our current results could be carried out by deriving explicit formulae for the passage from unstructured to structured force networks,  which do not require polyhedral stress functions. 
A second modification of the procedure described in Sect. \ref{stress}
relates to the use of mesh-free interpolation schemes, such as, e.g., the local maximum-entropy approach presented in \cite{FraMar:2012}.
Finally, another relevant generalization of the present research regards the prediction of the stress fields associated with fully 3D force networks.
In principle, such a challenging  extension might be accomplished by making use of Maxwell or Morera stress functions \cite{Sadd}, and applying the present procedures in correspondence with three different planes. 
However, the application of this approach to the development of provably convergent numerical
schemes for 3D stress field remains at present an open question, which we look forward to analyze in future studies.

\section*{ACKNOWLEDGEMENTS}
\section*{Acknowledgements}
Support for this work was received from the
Italian Ministry of Foreign Affairs, Grant No. 00173/2014, Italy-USA  Scientific and Technological Cooperation 2014-2015
(`\textit{Lavoro realizzato con il contributo del Ministero degli Affari Esteri, Direzione Generale per la Promozione del Sistema Paese}').
The authors would like
to thank Giuseppe Rocchetta (University of Salerno) and Robert Skelton (University of California, San Diego) for many useful discussions and suggestions.

\bibliographystyle{plain}

\begin{thebibliography}{}



\bibitem[O`Dwyer, 1999]{Dwy:1999}
O`Dwyer, D.,
Funicular analysis of masonry vaults, 
{\it Comput Struct}, 1999, 73, 187--197.

\bibitem[Fraternali, 2001]{LSM1}
Fraternali, F., 
Complementary energy variational approach for plane elastic problems with singularities,
{\it Theor. Appl. Fract. Mech.}, 2001, 35, 129--135

\bibitem[Fraternali et al., 2002]{LSM2}
Fraternali, F., Angelillo, M. and Fortunato, A., 
A lumped stress method for plane elastic problems and the discrete-continuum approximation,
{\it Int. J. Solids Struct.}, 2002, 39, 6211--6240.

\bibitem[Davini and Paroni, 2003]{DaviniParoni:2003}
Davini, C., and Paroni, R., 
\newblock{Generalized hessian and external approximations in variational problems of second order}, 
{\it J. Elast.}, 2003, 70, 149--174.

\bibitem[Kilian and Ochsendorf, 2005]{Kil:2005}
Kilian, A. and Ochsendorf, J.,
Particle-spring systems for structural form finding,
{\it IASS J.}, 2005, 46(2), 77--85

\bibitem[Block and Ochsendorf, 2007]{Blo:2007}
Block, P. and Ochsendorf, J.,
Thrust Network Analysis: A new methodology for three-dimensional equilibrium, 
{\it IASS J.}, 2007, 48(3), 167--173.

\bibitem[Block, 2009]{Blo:2009}
Block, P.,
Thrust Network Analysis: Exploring Three-dimensional equilibrium,
{\it Ph.D. dissertation, Massachusetts Institute of Technology, Cambridge, USA}, 2009.

\bibitem[Micheletti, 2008]{Micheletti:2008}
Micheletti, A.,
On generalized reciprocal diagrams for internally self-equilibrated frameworks,
{\it Int. J. Space Struct.}, 2008, 23, 153--166.

\bibitem[Fraternali, 2010]{LSM4}
Fraternali, F.,
A thrust network approach to the equilibrium problem of unreinforced masonry vaults via polyhedral stress functions,
 {\it Mech. Res. Commun.}, 2010, 37, 198--204.

\bibitem[Fraternali, 2011]{LSM5}
Fraternali, F.,
A mixed lumped stress--displacement approach to the elastic problem of masonry walls.
{\it Mech. Res. Commun.}, 2011, 38, 176--180

\bibitem[Block and Lachauero, 2011]{Blo:2011}
Block, P., and Lachauero, L.,
Closest-fit, compression-only solutions for free form shells, 
In: {\it IABSE/IASS London Sym-posium, Int. Assoc. Shell Spatial Structures}, 2011.

\bibitem[Angelillo et al., 2012]{Angelillo:2012}
Angelillo, M., Babilio, E., and Fortunato, A.,
Singular stress fields for masonry-like vaults,
{\it Continuum Mech. Therm.}, 2012, 25, 423-441.

\bibitem[Desbrun et al., 2013]{Desbrun:2013}
Desbrun, M., Donaldson, R., and Owhadi, H.,
\newblock{Modeling across scales: Discrete geometric structures in homogenization and inverse homogenization},
In: Multiscale analysis and nonlinear dynamics: from genes to the brain, Pesenson, M.Z., Ed., Vol. 8 of {\it Reviews of Nonlinear Dynamics and Complexity. Wiley}, 2013.

\bibitem[De Goes et al., 2013]{DeGoes:2013}
De Goes, F., Alliez, P., Owhadi, H. and Desbrun, M., 
\newblock{On the equilibrium of simplicial masonry structures},
{\it ACM Transactions on Graphics}, 2013, 32(4), 93.

\bibitem[Fraternali, 2007]{LSM3}
Fraternali, F., 
Error estimates for a lumped stress method for plane elastic problems,
{\it Mech. Adv. Matl. Struct.}, 2007, 14(4), 309--320.

\bibitem[Giaquinta and Giusti, 1985]{Giaquinta:1985}
Giaquinta, M., and Giusti, E.,
Researches on the equilibrium of masonry structures, 
{\it Arch. Ration. Mech. An.}, 1985, 88, 359--392.

\bibitem[Mansfield, 1964]{Mansfield:1964}
Mansfield, E.H.,
The Bending and Stretching of Plates,
{\it Pergamon Press}, 1964.

\bibitem[Miller and Tadmor, 2009]{Miller2009}
Miller, R. E., and Tadmor, E. B.,
A unified framework and performance benchmark of fourteen multiscale atomistic/continuum coupling methods, 
{\it J. Model. Simul. Mater. Sc.}, 2009, 17, 053001.

\bibitem[Liu et al., 2004]{Huang2004}
Liu, B., Huang, Y., Jiang, H., Qu, S., and Hwang, K. C.,
The atomic-scale finite element method,
{\it Comput. Methods Appl. Mech. Engrg.}, 2004, 193:1849--1864.

\bibitem[Tu and Ou-Yang, 2008]{Tu:2008}
Tu, Z.C., and Ou-Yang Z.C.,
Elastic theory of low-dimensional continua and its application in bio- and nano-structures,
{\it J. Comput. Theor.  Nanosci.}, 2008, 5, 422--448.

\bibitem[Fraternali et al., 2010]{Fra2:2010}
Fraternali, F., Blegsen, M., Amendola, A., and Daraio, C.,
Multiscale mass-spring models of carbon nanotube foams,
{\it J. Mech. Phys. Solid.}, 2010, 59(1), 89--102.

\bibitem[Raney et al., 2011]{CNTIDENT}
Raney, J.R., Fraternali, F., Amendola, A., and Daraio, C.,
{Modeling and in situ identification of material parameters for layered
structures based on carbon nanotube arrays},
{\it Compos. Struct.}, 2011, 93, 3013--3018.

\bibitem[Blesgen et al., 2012]{BFRAD11}
Blesgen, T., Fraternali, F.,  Raney, J.R., Amendola, A., and Daraio, C.,
{Continuum limits of bistable spring models of carbon nanotube arrays
accounting for material damage},
{\it Mech. Res. Commun.}, 2012, 45, 58--63. 

\bibitem[Fraternali et~al., 2012]{FraMar:2012}
Fraternali, F., Lorenz, C., and Marcelli, G.,
On the estimation of the curvatures and bending rigidity of membrane networks via a local maximum-entropy approach,
{\it J. Comput. Phys.}, 2012, 231, 528--540. 

\bibitem[Schmidt and Fraternali, 2012]{FraJMPS:2012}
Schmidt, B., and Fraternali, F.,
Universal formulae for the limiting elastic energy of membrane networks,
{\it J. Mech. Phys. Solids}, 2012, 60, 172--180.

\bibitem[Fraternali and Marcelli, 2012]{FraMar2:2012}
Fraternali, F., and Marcelli, G.,
A multiscale approach to the elastic moduli of biomembrane networks,
{\it Biomech Model Mechanobiol}, 2012, 11, 1097--1108.

\bibitem[Skelton, 2002]{Ske:2002}
Skelton, R.E.,
Structural systems: a marriage of structural engineering ans system science,
{\it J. Struct. Control}, 2002, 9, 113--133.

\bibitem[Vera et al., 2005]{Ver:2005}
Vera, C., Skelton, R.E., Bosscns, F., Sung, L.A.,
3-D nanomechanics of an erythrocyte junctional complex in equibiaxial and anisotropic deformations,
{\it Ann. Biomed. Eng.}, 2005, 33(10), 1387--1404. 

\bibitem[Mofrad and Kamm, 2006]{Mof:2006}
Mofrad, M.R.K., and Kamm, R.D.,
(Eds.), Cytoskeletal Mechanics: Models and Measurements,
{\it Cambridge University Press}, 2006.

\bibitem[Skelton and de Oliveira, 2010]{Ske:2010}
Skelton, R.E., and de Oliveira M.C.,
Tensegrity Systems,
{\it Springer}, 2010.

\bibitem[Fraternali et al., 2012]{Fra:2012}
Fraternali, F., Senatore, L., and Daraio, C.,
Solitary waves on tensegrity lattices,
{\it J. Mech. Phys. Solids}, 2012, 60, 1137--1144. 

\bibitem[Linhard and Bletzinger, 2010]{Lin:2010}
Linhard, J., and Bletzinger, K.--U.,
"Tracing" the Equilibrium -- Recent Advances in Numerical Form Finding, 
{\it International Journal of Space Structures}, 2010, 25(2), 107--116. 

\bibitem[Ohmori, 2011]{Ohmori:2011}
Ohmori, H.,
Computational Morphogenesis: Its Current State and Possibility for the Future, 
{\it International Journal of Space Structures}, 2011, 26(3), 269--276. 

\bibitem[Wang and Ohmori, 2010]{Wang:2010}
Wang, H., and Ohmori, H.,
Truss optimization using genetic algorithm, considering construction process, 
{\it International Journal of Space Structures}, 2010, 25(4), 202--215.

\bibitem[Baldassini et al., 2010]{Baldassini:2010}
Baldassini, N., Pottmann, H., Raynaud, J., and Schiftner, A.,
New strategies and developments in transparent free-form design: From facetted to nearly smooth envelopes,
{\it International Journal of Space Structures}, 2010, 25(3), 185--197.

\bibitem[Schlaich et al., 1987]{Schlaich}
Schlaich, J., Sch\"{a}fer, K.,  and Jennewein, M.,
Toward a Consistent Design  of Structural Concrete.
{\it Journal of Prestressed Concrete Institute (PCIJ)}, 1987 , 32, 74--150.

\bibitem[Shen and Atluri, 2004]{Shen2004}
Shen, S. and Atluri, S.N.,
Atomic-level stress calculation and continuum-molecular system equivalence,
{\it CMES - Comp Model. Eng.}, 2004, 6, 91--104. 

\bibitem[Admal and Tadmor, 2010]{Admal2010}
Admal, N. C., and Tadmor, E. B.,
A unified interpretation of stress in molecular systems,
{\it J. Elast.}, 2010, 100, 63--143

\bibitem[Glowinski, 1973]{Glowinski:1973}
Glowinski, R.,
{\it Approximations Externes, par 
Eelements Finis de Lagrange d'Ordre Un en Deux, du Probl\'{e}me de Dirichelet pour
l'Operateur Biharmonique. Methodes Iteratives de Resolution des Problemes Approches},
In: Miller, J.J.H. (Ed.), Topics in Numerical Analysis. Academic Press, 123--171, 1973.

\bibitem[De Guzm\'{a}n and Orden, 2006]{DeGuz:2006}
De Guzm\'{a}n, M., and Orden, D.,
\newblock{From graphs to tensegrity structures: Geometric and symbolic approaches},
{\it Publ. Mat}, 2006, 50, 279--299.

\bibitem[Strang, 2009]{Strang}
Strang, G.,
{\it Introduction to Linear Algebra},
4th Edition, Cambridge University Press, 2009.

\bibitem[Gurtin, 1972]{Gurtin}
Gurtin, M. E., 
{\it The Linear Theory of Elasticity},
In {Handbuch der Physik (Encyclopedia of Physics)}. VIa/2, Springer-Verlag, 1--295, 1972.

\bibitem[Green and Zerna, 2002]{Green:2002}
Green, A., and Zerna, W.,
Theoretical Elasticity, Dover, 2002.

\bibitem[Fraternali et~al., 2013]{Fracur:2013}
Fraternali, F., Farina, I., and Carpentieri, G.,
A discrete-to-continuum approach to the curvature of membrane networks and parametric surfaces,
{\it Mech. Res. Commun., in press}, 2013.

\bibitem[Cyron et al., 2009]{Cyr:2009}
Cyron, C.J., Arrojo, M., and Ortiz, M.,
Smooth, second-order, non-negative meshfree approximants selected by maximum entropy,
{\it Int. J. Num. Meth. Eng.}, 2009, 79, 1605--1632. 

\bibitem[Michell, 1904]{Michell}
Michell, A.G.M., 
The Limits of Economy of Materials in Frame Structures,
{\it Philos. Mag., Series 6}, 1904, 8(47), 589--597.

\bibitem[Akima, 1978]{Akima:1978}
Akima, H., and Ortiz, M.,
Algorithm 526: Bivariate Interpolation and Smooth Surface Fitting for Irregularly Distributed Data Points [E1],
{\it ACM T. Math. Software}, 1978, 4, 160--176.

\bibitem[Baratta and Corbi, 2010]{Baratta:2010}
Baratta, A., and Corbi, O.,
On the equilibrium and admissibility coupling in NT vaults of general shape,
{\it Int. J. Solids Struct.}, 2010, 47, 2276--2284.

\bibitem[Hegemier and Prager, 1969]{Prager:Michell}
Hegemier, G.A., and Prager, W.
On Michell trusses,
{\it Int. J. Mech. Sci.}, 1969, 11, 209--215.


\bibitem[Sadd, 2005]{Sadd}
Sadd, M. H., 2005. 
{\it Elasticity: Theory, Applications, and Numerics}, Elsevier.


\end{thebibliography}

\end{document}